\documentclass[a4paper,fleqn]{cas-sc}
\usepackage{algorithm}
\usepackage{algpseudocode}
\usepackage{amsmath} 
\usepackage{amsfonts} 
\usepackage{amssymb} 
\usepackage{booktabs}   
\usepackage{comment}
\usepackage{siunitx}    
\usepackage{caption}
\usepackage[numbers]{natbib}
\usepackage{color}
\usepackage{enumitem}
\usepackage{graphicx}   
\usepackage{subcaption}
\usepackage{multirow}   
\usepackage{tabularx}
\usepackage{ragged2e}   
\usepackage{rotating}
\usepackage{lscape} 
\usepackage{longtable}
\usepackage{svg}
\usepackage{float}

\usepackage{makecell} 
\usepackage{array}    

\def\tsc#1{\csdef{#1}{\textsc{\lowercase{#1}}\xspace}}
\tsc{WGM}
\tsc{QE}
\tsc{EP}
\tsc{PMS}
\tsc{BEC}
\tsc{DE}

\begin{document}
\let\WriteBookmarks\relax
\def\floatpagepagefraction{1}
\def\textpagefraction{.001}
\shorttitle{VSLLaVA}

\shortauthors{Qi Li et~al.} 

\title [mode = title]{VSLLaVA: a pipeline of large multimodal foundation model for industrial vibration signal analysis}       





%
\author[1,2]{Qi Li}[orcid = 0000-0001-7105-2818]
\ead{liq22@tsinghua.org.cn}

\author[1]{Xinran Zhang}[]
\ead{xinran-z24@mails.tsinghua.edu.cn}
    
\author[1]{Jinfeng Huang}[]
\ead{hjinfeng1991@163.com}

\author[1]{Hongliang He}[]
\ead{danson127hhl@gmail.com}

\author[1]{Feibin Zhang}[]
\cormark[1]
\ead{zfbin2008@163.com}



\author[1]{Zhaoye Qin}[orcid = 0000-0003-3892-4594]
\cormark[1]
\ead{qinzy@mail.tsinghua.edu.cn}

\affiliation[1]{organization={State Key Laboratory of Tribology, Department of Mechanical Engineering, Tsinghua University},
    city={Beijing},
    postcode={100084}, 
    country={P.R. China}}

\affiliation[2]{organization={Department of Statistics and Data Science, Yale University},
    city={New Haven},
    postcode={CT 06511}, 
    country={USA}}

\author[1]{Fulei Chu}[]
\ead{chufl@mail.tsinghua.edu.cn}

\cortext[cor1]{Corresponding author: Feibin Zhang, Zhaoye Qin}
\cortext[cor2]{Qi Li and Xinran Zhang contributed equally}

\begin{abstract}        
While Large Multimodal Models (LMMs) excel in general multimodal tasks, they lack the domain-specific knowledge for industrial vibration signal analysis. This paper introduces VSLLaVA, a comprehensive pipeline that utilizes expert knowledge-guided instruction tuning and evaluation to create an end-to-end LMM for signal analysis. To achieve this, we construct a novel Signal-Question-Answer (SQA) dataset using an expert rule-based signal generator. This dataset facilitates a two-stage learning procedure. The first step is efficient instruction fine-tuning with Low-Rank Adaptation (LoRA), which imparts specialized signal identification capabilities. Subsequently, we designed a tailored Group Relative Policy Optimization (GRPO) to refine the reasoning capabilities and enhance classification robustness. Then, a dual-mode evaluation framework is proposed, combining an LLM referee with expert rules for semantic assessment using quantitative metrics for numerical and textual accuracy, which reveals that VSLLaVA significantly improves performance in signal type identification and parameter analysis, and makes progress in the identification and parameter analysis of fault-related signals. This research demonstrates a viable approach for developing specialized foundational models for complex industrial applications and marks a transition from conventional task-specific systems to a cohesive, interactive foundational model.

\end{abstract}





\maketitle
\section{Introduction}          
\label{introduction}

Prognostics and Health Management (PHM) is a critical discipline for ensuring the reliability and safety of industrial systems by facilitating a shift from reactive to proactive, condition-based maintenance \cite{zio_prognostics_2022,kumar2023review}. This methodology encompasses a systematic process of data acquisition, anomaly detection, fault diagnosis, and remaining useful life prediction, which collectively inform maintenance decisions \cite{hu2022prognostics}. By doing so, PHM effectively reduces unexpected failures and operational costs, thereby extending the service life of machinery \cite{xu2021machine}.

The core of PHM is signal processing, which extracts condition-related information from sensor data \cite{althubaiti2022fault,lv2022vibration}. Traditional methods like Short-Time Fourier Transform \cite{gabor1946theory}, Wavelet Transform \cite{haar1909theorie}, and Hilbert–Huang Transform \cite{huang1998empirical} are vital for analyzing non-stationary signals from incipient faults. However, these methods come with constraints as they require in-depth domain knowledge for feature development, face challenges when dealing with intricate signals, and their labor-intensive process of manual feature engineering hinders practical effectiveness.
To overcome these challenges, AI-driven approaches using Machine Learning (ML) and Deep Learning (DL) have become transformative, automating feature extraction from data \cite{polverino2023machine,zhao2021challenges,thoppil2021deep}. Yet, their performance hinges on large, high-quality labeled datasets. This requirement poses a significant bottleneck in industrial settings, where fault data is inherently scarce and imbalanced, thus hindering model training and generalization.

More recently, the paradigm has shifted towards Large-Scale Foundation Models (LSFMs), such as Large Language Models (LLMs) and Large Multimodal Models (LMMs), offering new potential for PHM through their emergent reasoning and generalization capabilities \cite{li_chatgpt-like_2024, liu2024survey,alsaif2024multimodal,chen2025large}. Among these, LMMs excel at various instruction-guided tasks \cite{zhang2024vision,carolan2024review}. However, their application to PHM is impeded by fundamental challenges. The primary issues include:

\begin{itemize}
    \item \textbf{Absence of Domain Priors:} A primary limitation of current LMMs is the absence of inherent domain-specific priors for signal processing. Consequently, their outputs may violate underlying physical principles and operational constraints, especially in zero-shot or few-shot settings, which limits their effectiveness for industrial signal interpretation and fault diagnosis \cite{lai2024bearingfm}.

    \item \textbf{Fragmented and Inflexible Systems:} The prevailing paradigm in PHM relies on a fragmented ecosystem of specialized, multi-stage systems. These models are engineered for narrow tasks and lack a unified, interactive interface. Task guidance is hard-coded, making them inflexible and hindering the dynamic, human-in-the-loop analysis required in modern industrial settings \cite{hu2022prognostics}.

    \item \textbf{Modality and Semantic Gaps:} A persistent challenge is the modality and semantic gap between time-series signals and natural language. The heterogeneity between continuous, high-dimensional numerical representations and discrete, symbolic text hampers alignment; closing this representational disparity is essential to enable language to serve as an effective interface for signal interpretation \cite{wang2025innovative}.
\end{itemize}

To address these challenges, we introduce VSLLaVA, an end-to-end, instruction-tuned model designed to serve as a versatile, general-purpose tool that can be controlled and queried through natural language. We argue for and propose a paradigm shift: moving from building specialized, multi-stage systems to developing a single, unified, and interactive foundation model for industrial signal analysis. Our primary contributions are:

\begin{itemize}
    \item We present VSLLaVA, a domain-tailored pipeline that equips LMMs with expert priors for vibration signal analysis, yielding measurable improvements in signal identification and parameter analysis of fault-related signals.
    
    \item We construct an expert-guided Signal-Question-Answer (SQA) dataset from both simulated and real signals to enable multimodal instruction tuning.
    
    \item We perform a two-stage tuning process, including Low-Rank Adaptation (LoRA) to align signal and language modalities, and tailored Group Relative Policy Optimization (GRPO) with a task-specific composite reward, to improve reasoning and enhance robustness in signal identification.

    \item We introduce a dual-mode evaluation framework that combines an automated LLM referee with expert assessment for comprehensive, reproducible validation.
    
\end{itemize}

The rest of this paper is organized as follows. Section \ref{related_work} provides related works about signal analysis and LMMs in PHM. Section \ref{method} presents the proposed VSLLaVA pipeline. Section \ref{experiments} describes the experimental setup, results, and discussion. Section \ref{conclusion} presents the conclusion and outlines future work. The notations can be seen in Table \ref{tab:notations_comprehensive}.

\begin{table}[htbp ]
\centering
\caption{Nomenclature}
\label{tab:nomenclature}
\begin{tabular}{ll}
\toprule
\textbf{Abbreviation} & \textbf{Full Term}  \\
\midrule
PHM          & Prognostics and Health Management     \\
ML           & Machine Learning                      \\
DL           & Deep Learning                         \\
LLM          & Large Language Model                  \\
LMM          & Large Multimodal Model                \\
LSFM         & Large-Scale Foundation Model          \\
LoRA         & Low-Rank Adaptation                   \\
GRPO         & Group Relative Policy Optimization    \\
SQA          & Signal-Question-Answer                \\
QA           & Question-Answer                       \\
SFT          & Supervised Fine-Tuning                \\
\bottomrule
\end{tabular}
\end{table}


\begin{table*}[htbp ]

\centering
\caption{Notations used throughout Section \ref{method}.}
\label{tab:notations_comprehensive}
\renewcommand{\arraystretch}{1.2}
\setlength{\tabcolsep}{4pt} 

\begin{tabularx}{\textwidth}{@{} l X @{}}
\toprule
\textbf{Symbol} & \textbf{Definition} \\
\midrule

\multicolumn{2}{l}{\textbf{Section \ref{SFT}: LMM and Fine-Tuning}} \\
\midrule
$I_s$ & Input signal image \\
\( P_{\Phi_e}(Z_s|X_s) \) & The pretrained encoder in an LVM. \\
\( P_{\Phi_a}(X_a|H_q) \circ P_{\Phi_q}(H_q|X_q) \) & The pretrained LLM in an LVM. \\
\( P_{\Phi_m}(H_s|Z_s) \) & The modal alignment unit in an LVM. \\
$G$ & The SQA group of a signal. \\
$\mathbf{X}_a, X_a$ & The answer sequence and a piece of answer. \\ 
$\mathbf{X}_q, X_q$ & The question sequence and a piece of question. \\
$\mathbf{X}_p, X_p$ & The model prediction sequence and a piece of prediction. \\
$\theta, \theta_0$ & All trainable model parameters and the original parameters of the LLM. \\
$\Delta\theta(\Theta), \Theta $ & Parameter increment in LoRA and the smaller set of parameters defining the increment. \\
$A, B$ & Low-rank matrices for LoRA where $\Delta W = AB$. \\
$r$ & The rank of the LoRA decomposition. \\
\midrule

\multicolumn{2}{l}{\textbf{Section \ref{GRPO}: GRPO Reward Function}} \\
\midrule
$\mathbf{X}_c, X_c$ & The sequence of candidate model completions and a piece of completion. \\
$R$ & The rewards list. \\
$a, l$ & A single model-generated answer string and a true label string. \\
$\mathcal{V}, \mathcal{A}_t$ & Pre-defined weighted synonym vocabulary and the set of acceptable answers for a label. \\
$\sigma, \sigma_{\text{best}}$ & Fuzzy matching score and the best Fuzzy matching score. \\
$w, \omega$ & A synonym string and its corresponding professionality weight. \\
$w_{\text{best}}, \omega_{\text{best}}$ & The best matching synonym string and its corresponding professionality weight. \\
$S_{\text{reward}}$ & Calculated reward score. \\
$\beta_{\text{exact}}$ & Exact Match Bonus. In this work, we set $\beta_{\text{exact}}=0.1$. \\
\midrule

\multicolumn{2}{l}{\textbf{Section \ref{evaluation_method}: Evaluation Metrics}} \\
\midrule
$S_n$ & The Numerical Score. \\
$S_w$ & The Word Recall. \\
$S_\text{BLEU-1}, S_\text{BLEU-2}, S_\text{BLEU-3}, S_\text{BLEU-4}$ & The score of BLEU-1, BLEU-2, BLEU-3, BLEU-4 respectively. \\
$S_\text{ROUGE-1}, S_\text{ROUGE-2}, S_\text{ROUGE-l}$ & The score of ROUGE-1, ROUGE-2, ROUGE-l respectively. \\
$S_\text{CIDEr}$ & The score of CIDEr.\\
$\mathbf{v}_{\text{ref}}, \mathbf{v}_{\text{pred}}$ & The sequence of numbers extracted from the standard answer and the model prediction. \\
$\mathcal{W}_{\text{ref}}, \mathcal{W}_{\text{pred}}$ & The set of unique words in the standard answer and the model prediction. \\
$\lambda$ & The hyperparameter representing the constraint of the mean relative error. In this paper, we set $\lambda=1$. \\
$\mathcal{D}, d$ & The whole SQA evaluation dataset and an SQA sample. \\
$C$ & The referee model configuration file. \\
$R_\text{llm}, R_\text{custom}, R_\text{analysis}$ & The referee LLM evaluation results, the custom metrics evaluation results and the specialized analysis results. \\
$S_\text{llm}$ & The score presented by the referee model. \\
$S_\text{avg\_group}$ & The score containing the average value of each metric for each SQA group of different signal types. \\
$S_\text{avg\_overall}$ & The score containing the average score of each metric for the entire dataset with each group macro-averaged. \\

\bottomrule
\end{tabularx}
\end{table*}

\clearpage

\section{Related work} 
\label{related_work}

\subsection{Industrial vibration signal analysis} 

Industrial vibration signal analysis is a crucial technique for PHM, involving signal pre-processing and feature extraction for tasks such as anomaly detection and fault diagnosis \cite{Lee2014}. Classic methods like envelope analysis \cite{Randall2011} and the Wavelet Transform \cite{haar1909theorie} excel at identifying characteristic frequencies in non-stationary signals, aiding feature extraction. More recently, ML and DL have been introduced to automate the extraction of rich, discriminative features. For instance, various studies have employed diverse neural network architectures, including those combining Discrete Wavelet Transform with feature selection classifiers \cite{hosseinpour-zarnaq_fault_2022}, multi-head convolutional neural networks for multi-channel signals \cite{junior_fault_2022}, and multi-input models that leveraged multi-dimensional signal features \cite{wang_multi-input_2021}. Other advanced approaches focused on learning domain-invariant features to handle varied operating conditions \cite{fan2023deep}, improving feature learning via specialized morphological filtering layers \cite{ye_deep_2021}, or addressing data scarcity and domain shift through fault-aware autoencoders and contrastive learning strategies \cite{pang2024fault}.

Despite their demonstrated efficacy, these methods are often task-specific and lack general-purpose interactivity. They typically function as end-to-end systems that, once trained, cannot dynamically adapt to novel queries beyond their predefined scope. This operational rigidity contrasts sharply with the paradigm offered by LSFMs, which can reason over diverse data modalities in a zero-shot or few-shot context. Unlike traditional methods, LSFMs support the interactive, exploratory, and multifaceted analytical needs of modern industrial environments. This highlights a compelling research gap: the need for a versatile large model tailored for industrial signal processing, one that integrates the pattern recognition strengths of DL with the general-purpose, interactive reasoning of LSFMs.

\subsection{Large-scale foundation models in PHM} 

Recent years have witnessed a paradigm shift towards large-scale, pre-trained foundation models. While large language models have shown remarkable emergent capabilities, their single-modality nature constrains their application in complex signal analysis \cite{Fan2024}. To address this, LMMs have been developed, with large vision-language models like LLaVA \cite{liu2023llava} and DeepSeek-VL \cite{deepseekai2025deepseekr1incentivizingreasoningcapability,lu2024deepseekvlrealworldvisionlanguageunderstanding} gaining significant attention. Consequently, research has begun exploring the potential of applying LLMs to address long-standing challenges in PHM, generally focusing on two main directions: improving training data and enhancing model architecture.

One line of research aims to improve LMM performance by enhancing data quality. This includes fine-tuning models on industrial texts like technical documents and maintenance logs \cite{jose2024advancing}, establishing prompt-based frameworks to build few-shot learning capabilities \cite{zheng2024empirical}, and proposing foundation models that fuse signal and language modalities for fault diagnosis and RUL prediction \cite{wang2025innovative}. 
A complementary line of research focuses on architectural advances. Representative efforts include integrating domain knowledge and contrastive learning within semi-supervised frameworks to improve generalization \cite{lai2024bearingfm}; developing multimodal pipelines that couple an LLM with a specialized fault-classification network using prior-knowledge–enhanced signal representations \cite{peng2025bearllm}; employing fuzzy semantic embeddings to mitigate pattern confusion in feature spaces \cite{lin2025fd}; and introducing regression frameworks that leverage transfer learning to capture complex temporal dependencies in sensor data \cite{chen2024remaining}.

Building on these advances with LLM for specific tasks, we present VSLLaVA, a pipeline to build an LMM for signal analysis and fault diagnosis by jointly leveraging textual and signal inputs. We construct a large-scale signal dataset and fine-tune VSLLaVA on custom SQA triplets using LoRA. Performance is evaluated with a dual-mode framework that combines an external referee LLM with expert review. We further apply a tailored GRPO with a task-specific reward to enhance signal identification. Experiments demonstrate significant improvements in signal analysis and parameter identification.

\section{Method}                
\label{method}

\subsection{Overall framework}        

\begin{figure*}[htbp]
\centerline{\includegraphics[width=1\linewidth]{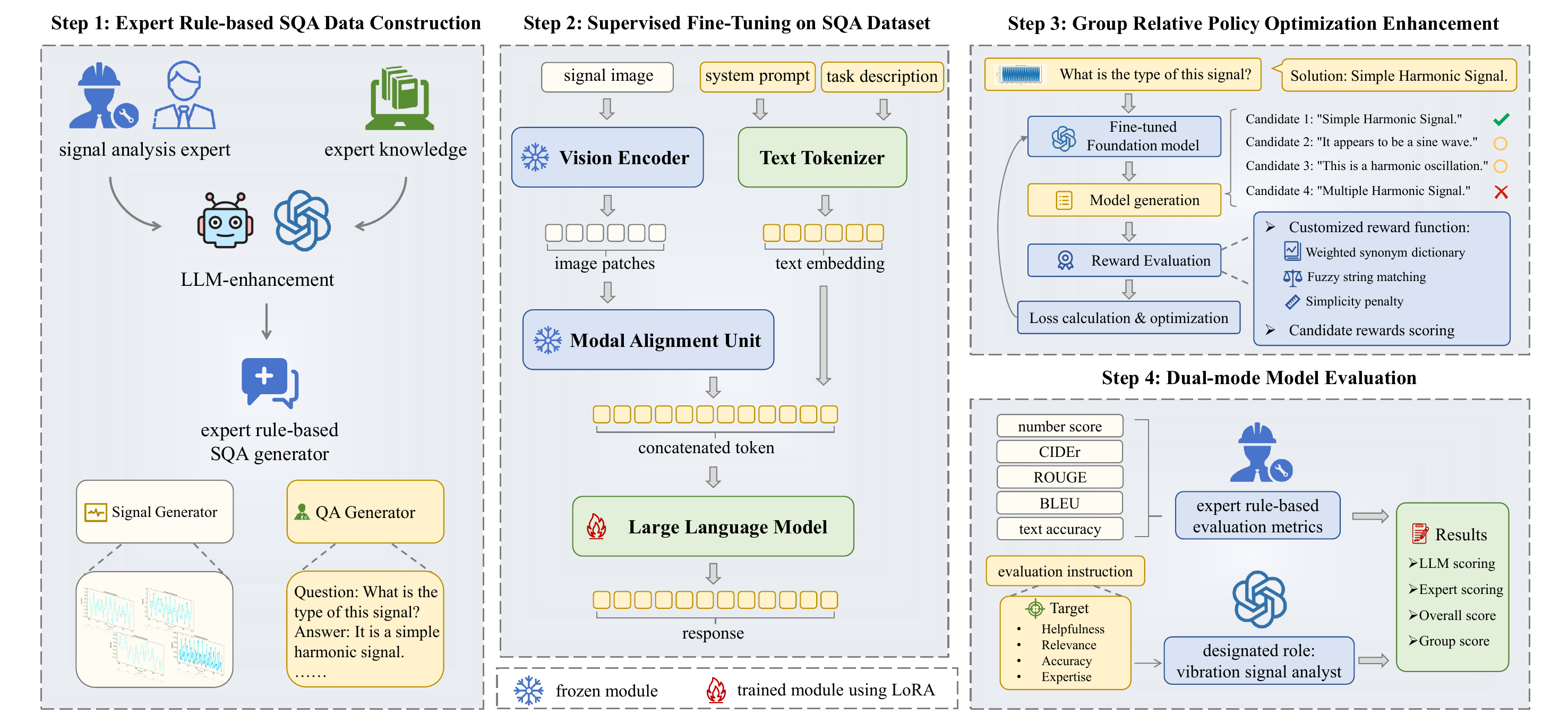}}
\caption{Pipeline of VSLLaVA.}
\label{fig framework}
\end{figure*}

To bridge the gap between LMMs and industrial vibration signal analysis, we propose VSLLaVA as shown in Fig. \ref{fig framework}, a pipeline of large multimodal foundation model enhanced with domain-specific adaptations. We used InternVL3-8B as the base model and applied LoRA techniques to fine-tune the linear layers of the language model. The fine-tuned model was then evaluated in collaboration with the LLM signal experts to assess the accuracy and relevance of the responses. In addition, to measure the reasoning abilities of VSLLaVA, we adopted GRPO to the fine-tuned model with a tailored reward function for model evaluation. 

\subsection{Expert rule-based signal generator}     

A significant challenge in applying LMMs to signal processing is the scarcity of specialized signal-text paired data, which is essential for imbuing these models with the requisite domain knowledge for vibration analysis. To address this gap, we developed a systematic methodology for generating SQA triplets using a suite of expert rule-based signal generators. These generators are designed to produce a comprehensive dataset encompassing a wide range of signal types, from fundamental waveforms to complex compositions, as detailed in Table \ref{signal generators}.

The construction of our SQA dataset is guided by the practical needs of industrial signal analysis, where specific signal parameters are critical for diagnosing mechanical faults. We generated SQA triplets for eleven foundational signal types, including various modulated, harmonic, and impulse signals. To incorporate real-world complexity, we augmented this synthetic data with the THU dataset, an experimental dataset capturing vibration signals in the voltage modality from rolling bearings under four distinct health conditions (normal, inner race fault, ball fault, and outer race fault) sampled at 49.6 kHz \cite{Zhang2022}. For each of the twelve signal categories (eleven synthetic and one real), our generators produce structured SQA triplets. These triplets encapsulate key domain knowledge, covering: 1) fundamental descriptions of the signal and signal parameters; 2) time- and frequency-domain characteristics, such as amplitude and peak frequencies; and 3) task-specific diagnostic assessments, particularly for the THU fault data. Visual representations of these signals are provided in Table \ref{Signal visualization} in the Appendix \ref{app:vis_data}.

\begin{table*}[h!]
\centering
\caption{Parameters for the expert rule-based signal generators. For modulated signals, parameters include a shared carrier frequency \( f_c \) and modulation frequency \( f_m \), max frequency deviation \( \Delta f \), and modulation index \( m \). For other signal types, parameters include amplitude \( A_i \), base frequency \( f_b \), random frequencies \( f_i \), phase angle \( \phi_i \), decay coefficient \( \beta \), and period or time offset \( T_0 \). The unit of frequency is uniformly defined as Hertz (Hz); the unit of amplitude is uniformly defined as Voltage (V); the unit of phase is uniformly defined as radians (rad).}
\label{signal generators}

\small 
\setlength{\tabcolsep}{4pt} 
\renewcommand{\arraystretch}{1.75} 

\begin{tabular}{c c c c}
\hline
\textbf{Type} & \textbf{Dataset} & \textbf{Equation}& \textbf{\makecell{Identify\\Parameters}} \\
\hline
\multirow{3}{*}{\makecell{Modulated\\Signal}} 
& \makecell{\textbf{A}mplitude \textbf{M}odulated} & $y_{AM}=\left[1+m\cos(2\pi f_m t)\right]\cdot\cos(2\pi f_c t)$ & $m, f_c, f_m$ \\
& \makecell{\textbf{F}requency \textbf{M}odulated} & $y_{FM}=\cos \left[2\pi f_c t + \frac{\Delta f}{f_m} \cdot \sin(2\pi f_m t)\right]$ & $\Delta f, f_c, f_m$ \\
& \makecell{Composite \textbf{AM-FM}} & $y_{AMFM} = y_{AM} + y_{FM} $ & $m, f_c, f_m, \Delta f$\\

\hline

\multirow{4}{*}{\makecell{Sinusoidal\\Signal}} 
& \makecell{\textbf{S}imple \textbf{H}armonic} & $y_{SH} =A \sin(2 \pi f_b t + \phi) $ & $A, f_b, \phi$ \\
& \makecell{\textbf{M}ultiple \textbf{H}armonic} & $ y_{MH}=\sum_k A_{k}\sin(2\pi k f_b t+\phi_{k})$ & $\{A_k\}, f_b, \{\phi_k\}$ \\
& \makecell{\textbf{R}andom \textbf{H}armonic} & $ y_{RH}=\sum_j A_{j}\sin(2\pi f_j t+\phi_{j}), \text{ where } f_j \sim \mathcal{N}(\mu_f, \sigma_f^2)$ & $\{A_j, f_j, \phi_j\}$\\
& \makecell{\textbf{C}omposite \textbf{H}armonic} & $y_{CH}=y_{MH} + y_{RH}$ & 
$\makecell{f_b, \{A_k, \phi_k\}_{\text{MH}}, \{A_j, f_j, \phi_j\}_{\text{RH}}}$\\

\hline

\multirow{4}{*}{\makecell{Impulse\\Signal}} 
& \makecell{\textbf{S}ingle \textbf{T}ransient} & $y_{ST} = A e^{-\beta t} \cdot \sin(2\pi f_b t + \phi) \cdot u(t)$ & $A, \beta, f_b, \phi $ \\
& \makecell{\textbf{M}ultiple \textbf{T}ransient} & $y_{MT} = \sum_i A_i e^{-\beta_i t } \cdot \sin(2\pi f_i t + \phi_i) \cdot u_i(t) $ & $\{A_i, \beta_i, f_i, \phi_i\}$ \\
& \makecell{\textbf{S}ingle \textbf{P}eriodic} & $y_{SP} = A e^{-\beta(t - T_0)} \cdot \sin[2\pi f_b(t - T_0) + \phi]$ & $A, \beta, T_0, f_b, \phi$ \\
& \makecell{\textbf{M}ultiple \textbf{P}eriodic} & $y_{MP} = \sum_i A_i e^{-\beta(t - i T_0)} \cdot \sin[2\pi f_b(t - i T_0) + \phi_i]$ & \makecell{$\{A_i\}, \beta, T_0, f_b, \{\phi_i\}$} \\
\hline 
Real Signal & \textbf{THU} Signal & / & / \\
\hline
\end{tabular}
\end{table*}

\subsection{Multimodal model tuning with SQA data} 
\label{SFT}

The core learning capability of VSLLaVA resides in adapting a pre-trained LMM to the specialized domain of vibration signal analysis. Our approach follows the widely-adopted "connector-based" LMM paradigm composed of three main components: a pre-trained vision encoder \( P_{\Phi_e}(Z_s|X_s) \), a language model \( P_{\Phi_a}(X_a|H_q) \circ P_{\Phi_q}(H_q|X_q) \), and a lightweight modal alignment unit \( P_{\Phi_m}(H_s|Z_s) \) that bridges them.

For each signal image $I_s$, we use an expert rule-based signal generator with domain knowledge to generate an SQA group \(G =  (I_s, X_{q1}, X_{a1}, \ldots, X_{qi}, X_{ai}, \ldots, X_{qN}, X_{aN}) \) to train the VSLLaVA to identify the parameters of the signal for the task, where \( N \) is the number of QA pairs. From this group, an instruction pair \( \mathbf{X_{instruct}} = (I_s, X_{q1}, X_{a1}) \) is extracted for the first turn, and pairs \( \mathbf{X_{instruct,i}} = (X_{qi}, X_{ai}) \) are used for the remaining turns.

Given $I_s$ as the input, the vision encoder first extracts a set of corresponding feature vectors. The modal alignment unit, typically a multi-layer perceptron, then projects these visual features into the word embedding space of the LLM, yielding language-compatible visual embeddings. These visual embeddings act as a soft prompt, enabling the LLM to "see" and reason about the image content. The entire process is trained end-to-end by maximizing the likelihood of generating the correct answer sequence $\mathbf{X}_a=(X_{a1}, X_{a2}, \ldots, X_{ai}, \ldots, X_{aN})$ given the signal image $I_s$ and the question $\mathbf{X}_q=(X_{q1}, X_{q2}, \ldots, X_{qi}, \ldots, X_{qN})$. The objective for a single SQA sample is to minimize the negative log-likelihood $\mathcal{P}(\theta)$, which can be denoted as

\begin{equation}
P(\theta) = -\log P(\mathbf{X}_a | I_s, \mathbf{X}_q) = -\sum_{i=1}^N \log P_\theta(X_{ai} | I_s, \mathbf{X}_{q,<i}, \mathbf{X}_{a,<i}),
\end{equation}
where $\mathbf{X}_{q,<i}$ and $\mathbf{X}_{a,<i}$ represents the preceding question and preceding answer up to the \( i \)-th step respectively, and $\theta$ denotes all trainable model parameters.

However, fine-tuning all parameters of a large model is computationally prohibitive and risks catastrophic forgetting. To address this, we employ a Parameter-Efficient Fine-Tuning (PEFT) technique called LoRA. The key insight of LoRA is that the change in weights for a pre-trained model during adaptation to a new task has a low intrinsic rank. Therefore, instead of directly updating the original weight matrix $W_0 \in \mathbb{R}^{d \times k}$ of a linear layer in the LLM, LoRA keeps $W_0$ frozen and introduces two trainable, low-rank matrices $A \in \mathbb{R}^{d \times r}$ and $B \in \mathbb{R}^{r \times k}$ to represent the weight update $\Delta W$. The rank $r$ is a small hyperparameter ($r \ll d, k$). The modified forward pass for a layer becomes:

\begin{equation}
\mathbf{Y} = \mathbf{X}W = \mathbf{X}(W_0 + \Delta W) = \mathbf{X}W_0 + \mathbf{X}AB,
\end{equation}
where $\mathbf{X}$ and $\mathbf{Y}$ represents the input and output of the linear layer in the LLM respectively. Therefore, LoRA can dramatically reduces the number of trainable parameters. 

Let the original parameters of the LLM be $\theta_0$, which are frozen. The fine-tuning process only optimizes a much smaller set of parameters that define the LoRA matrices $\{A, B\}$ for all adapted linear layers. The optimization objective over the entire SQA dataset can be denoted as follows

\begin{equation}
\max_{\Theta} \sum_{G} \sum_{i=1}^N \log P_{\theta_0 + \Delta\theta(\Theta)}(x_{a,i} | I_s, \mathbf{X}_{q,<i}, \mathbf{X}_{a,<i}),
\label{eq:lora_objective}
\end{equation}
where $\Delta\theta(\Theta)$ represents the task-specific parameter increment which is encoded by a much smaller set of parameters \( \Theta \), with \( |\Theta| \ll |\theta_{0}| \). The task of finding \( \Delta\theta \) becomes optimizing over \( \Theta \). In our VSLLaVA pipeline, we froze the vision encoder and the modal alignment unit and only trained the LLM, where LoRA was applied to all linear layers. This allows the model to efficiently learn the specific patterns and terminology of vibration signal analysis from our SQA triplets, ultimately producing a specialized yet robust diagnostic tool.

\subsection{Group Relative Policy Optimization in Signal Identification} 
\label{GRPO}

While fine-tuning with our SQA triplets provided VSLLaVA with foundational domain knowledge, this initial stage exhibited limitations characteristic of behavioral cloning. The model tended to replicate the verbose, explanatory style of the training data, producing lengthy responses rather than the concise, definitive classifications required for an expert system. Furthermore, the model performance was sensitive to phrasal variations, lacking the robustness needed for practical applications. The SFT phase produced a "knowledgeable collaborator," whereas the objective was to develop a "decisive expert."

To bridge this gap, we introduced a tailored GRPO \cite{shao2024deepseekmath} as a second-stage refinement strategy. The primary goal of this stage was to sharpen the model performance on the core task: precise and robust signal type identification. Unlike traditional reinforcement learning methods reliant on pairwise comparisons, GRPO is an advanced preference alignment algorithm which employs a group-level ranking mechanism. For a given input, the model generates a set of candidate answers, and the optimization objective is to maximize the log-likelihood of the best-ranked answer as determined by a reward function. This contrastive learning paradigm compels our VSLLaVA to distinguish between optimal and suboptimal responses, thereby refining its decision-making policy.

A generic reward model would be insufficient for the nuanced requirements of this domain. Therefore, we designed a custom reward function tailored specifically for signal identification as detailed in Algorithms \ref{alg:grpo_main} through \ref{alg:calculate_reward}, the notations of which is listed in Table \ref{tab:notations_comprehensive}. This function integrates three key mechanisms to guide the optimization process effectively:

\begin{enumerate}
    \item \textbf{Domain-Specific Semantic Mapping:} A weighted synonym vocabulary $\mathcal{V}$ maps various acceptable terms for signal types to a standard label. Each synonym $w$ is assigned a weight $\omega$ reflecting its technical precision.
    \item \textbf{Robust Fuzzy Matching:} To enhance robustness against minor spelling or syntactic variations, the reward function uses a fuzzy string matching algorithm to calculate the similarity between the model generation and the synonyms in the vocabulary. We represent our synonym vocabulary and corresponding weights in Appendix \ref{app:GRPO_details}. 
    \item \textbf{Incentive for Precision:} Instead of a penalty, a bonus term, $\beta_{\text{exact}}$, is introduced to encourage precision. This bonus is added to the reward score when the model's output perfectly matches one of the synonyms in the vocabulary ($\sigma_{\text{best}} = 100$), thereby incentivizing the model to generate the most ideal and concise answer.
\end{enumerate}

\begin{algorithm}[]
\caption{GRPO Reward Calculation (Main Process)}
\label{alg:grpo_main}
\begin{algorithmic}[1]
\Function{RewardCalculation}{the sequence of candidate model completions $\mathbf{X}_c$, the answer sequence $\mathbf{X}_a$, exact match bonus$\beta_{\text{exact}}$} \Comment{See Table \ref{tab:notations_comprehensive} for notations}
    \State \textbf{Initialize} rewards list $R$
    \For{$X_c$ in $\mathbf{X}_c$}
        \State Extract answer string $a$ from $X_c$
        \State Extract true label string $l$ from $X_a$
        \State Get acceptable answers $\mathcal{A}_t \gets \mathcal{V}[l_t]$
        \If{$\mathcal{A}_t$ is not defined}
            \State Append $0.0$ to $R$ and \textbf{continue}
        \EndIf
        \State $(\sigma_{\text{best}}, w_{\text{best}}, \omega_{\text{best}}) \gets \text{FindBestMatch}(a, \mathcal{A}_t)$ \Comment{See Alg. \ref{alg:find_best_match}}
        \State $S_{\text{reward}} \gets \text{CalculateReward}(a, \sigma_{\text{best}}, w_{\text{best}}, \omega_{\text{best}}, \beta_{\text{exact}})$ \Comment{See Alg. \ref{alg:calculate_reward}}
        \State Append $S_{\text{reward}}$ to $R$
    \EndFor
    \State \Return $R$
\EndFunction
\end{algorithmic}
\end{algorithm}

\begin{algorithm}[]
\caption{Helper Function: FindBestMatch}
\label{alg:find_best_match}
\begin{algorithmic}[1]
\Function{FindBestMatch}{A single model-generated answer string $a$, the set of acceptable answers for a label$\mathcal{A}_t$} 
    \State \Comment{See Table \ref{tab:notations_comprehensive} for notations}
    \State \textbf{Initialize} $\sigma_{\text{best}} \gets 0, w_{\text{best}} \gets \text{""}, \omega_{\text{best}} \gets 0.0$
    \For{each $(w, \omega)$ in $\mathcal{A}_t$}
        \State Calculate the $\sigma$ using $a, w$
        \If{$\sigma > \sigma_{\text{best}}$}
            \State $\sigma_{\text{best}} \gets \sigma$
            \State $w_{\text{best}} \gets w$
            \State $\omega_{\text{best}} \gets \omega$
        \EndIf
    \EndFor
    \State \Return $(\sigma_{\text{best}}, w_{\text{best}}, \omega_{\text{best}})$
\EndFunction
\end{algorithmic}
\end{algorithm}

\begin{algorithm}[]
\caption{Helper Function: CalculateReward}
\label{alg:calculate_reward}
\begin{algorithmic}[1]
\Function{CalculateReward}{$a$, the best Fuzzy matching score$\sigma_{\text{best}}$, the best matching synonym string$w_{\text{best}}$, corresponding professionality weight$\omega_{\text{best}}$, $\beta_{\text{exact}}$} \Comment{See Table \ref{tab:notations_comprehensive} for notations}
    \State Calculate the base reward directly from the fuzzy score and weight $S_{\text{reward}} \gets (\sigma_{\text{best}} / 100.0) \times \omega_{\text{best}}$
    \If{$\sigma_{\text{best}} = 100$}
        \State Add a bonus for a perfect match $S_{\text{reward}} \gets S_{\text{reward}} + \beta_{\text{exact}}$
    \EndIf
    \State $S_{\text{reward}} \gets \max(0, \min(1.0, S_{\text{reward}}))$
    
    \State \Return $S_{\text{reward}}$
\EndFunction
\end{algorithmic}
\end{algorithm}

In summary, through this two-stage paradigm of “SFT knowledge injection” and “GRPO strategy sharpening,” supplemented by a carefully designed customized reward function, we aimed to efficiently shape our VSLLaVA into a professional tool with high accuracy and robustness in industrial vibration signal recognition. The procedure of our GRPO experiments is displayed in Fig. \ref{fig framework}.

\subsection{Dual-mode model evaluation framework} 
\label{evaluation_method}

Following SFT and GPRO, assessing LMM poses notable difficulties \cite{Chang2024}. While automated evaluation provides a scalable means of assessing model outputs \cite{yang2022glue}, recent work has increasingly employed LLMs as referees to introduce a degree of semantic judgment and reduce human subjectivity \cite{chen2023exploring}.

To provide a robust and multifaceted assessment, this work introduces a dual-mode evaluation framework, which synergistically combines qualitative assessment from an automated LLM referee with human expert evaluation composed of a suite of objective, quantitative metrics. The LLM referee evaluates the logical coherence, semantic accuracy, and linguistic fluency of the generated responses. Concurrently, the human expert uses quantitative metrics to measure model performance across several dimensions, including numerical precision, textual similarity, and content consensus, thereby offering a comprehensive and reproducible overview of model capabilities.

\subsubsection{Multi-dimensional quantitative metrics} 

Specifically, in order to make a comprehensive evaluation of the number accuracy in parameter identification tasks and the semantic accuracy in signal recognition and description tasks, we used five evaluation metrics to evaluate the performance after fine-tuning. Among all, two metrics were self-customized to assess signal identification abilities of the model, including Numerical Score and Word Recall; the other three metrics directly adopted the existing benchmark metrics in natural language processing-the BLEU score \cite{papineni2002bleu}, the ROUGE score \cite{lin2004rouge} and the CIDEr score \cite{vedantam2015cider}-to measure the semantic accuracy of the model generations. In this section, we only discuss the customized metrics-Numerical Score and Word Recall metrics. 

\textbf{Numerical Score}. To evaluate the model performance on signal parameter identification tasks that require models to extract and calculate values from the signal images, we designed a scoring rule named Numerical Score which involves two core steps: calculating the mean relative error $E_{\text{mean}}$, and converting $E_{\text{mean}}$ into the final score $S_n$ using an exponential decay function. 
    
Let the sequence of numbers extracted from the standard answer be $\mathbf{v}_{\text{ref}}=(v_{\text{ref},1}, \dots, v_{\text{ref},n})$, and the sequence of numbers extracted from the model prediction be $\mathbf{v}_{\text{pred}}=(v_{\text{pred},1}, \dots, v_{\text{pred},k})$. The index of numbers to be compared is $k'=\min(n,k)$. First, calculate the relative error $E_i$ for the $i$-th pair of numbers $(v_{\text{ref},i}, v_{\text{pred},i})$ using Eq. \ref{eq:relative_error}:

\begin{equation}        
E_i = \frac{|v_{\text{pred},i}-v_{\text{ref},i}|}{\max(|v_{\text{ref},i}|,\epsilon)}.
\label{eq:relative_error}
\end{equation}

Note that for $|v_{\text{ref},i}|=0$, indicating that the standard answer does not contain numbers, we use $\epsilon=10^{-6}$ to ensure that the division makes sense. Then, calculate the average relative error $E_{\text{mean}}$ of these $k'$ pairs of numbers using Eq. \ref{eq:mean_relative_error}:

\begin{equation}        
E_{\text{mean}} = \frac{1}{k'} \sum_{i=1}^{k'} E_i.
\label{eq:mean_relative_error}
\end{equation}

Finally, calculate the final Numerical Score $S_n$ using an exponential decay function, which not only maps large, discrete errors to continuous fractions between $(0,1]$, but also aligns with the BLEU and ROUGE metrics in terms of range, providing a more intuitive indication of whether the model can accurately identify the parameters of unknown signals:

\begin{equation}            
S_n = \exp(- \lambda \cdot E_{\text{mean}}),
\label{eq:number_score}
\end{equation}
where $\lambda$ is the hyperparameter manually tuned by experts and represents the constraint of the $E_{\text{mean}}$. In this paper, we set $\lambda=1$. When the relative error is 0, the highest score is 1, and the larger the error, the closer the score is to 0. For scenarios where the standard answer does not contain numerical values or the model cannot provide numerical values, we have set $S_n$ to $\text{NAN}$ to indicate that this metric is invalid. In the process that calculates the average number score on each SQA dataset of different signals and on the entire SQA dataset, the results with a value of $\text{NAN}$ will be skipped, and the number of valid samples will be finally reported. Therefore, we utilized the Number Score to reflect the correctness of identified signal parameters. 

\textbf{Word Recall}. As a supplementary measure, we also calculated the recall of words between the predicted vocabulary set and the standard answer vocabulary set, which intuitively reflects the degree of keyword matching. It is calculated by using Eq. \ref{eq:word_recall}.

\begin{equation}            
    S_w = 
    \begin{cases} 
    \frac{|\mathcal{W}_{\text{ref}} \cap \mathcal{W}_{\text{pred}}|}{|\mathcal{W}_{\text{ref}}|} \times 100\% & \text{if } |\mathcal{W}_{\text{ref}}| \neq 0, \\
    100\% & \text{if } |\mathcal{W}_{\text{ref}}| = 0. 
    \end{cases}
    \label{eq:word_recall}
\end{equation}     

Note that $\mathcal{W}_{\text{ref}}$ is the set of unique words in the standard answer, while $\mathcal{W}_{\text{pred}}$ is the set of unique words in the model prediction. $S_w$ refers to the Word Recall, indicating what proportion of all words in the standard answers are covered by the model's predictions. During the process of model generations, we lowercased all tokens, stripped punctuation, removed stopwords, and lemmatized English words before forming $\mathcal{W}_{\text{ref}}$ and $\mathcal{W}_{\text{pred}}$. We also canonicalized units and numeric strings to ensure that the processed generations are as standardized as possible.

\textbf{BLEU}. To reflect the fluency of the text and the accuracy of word choice, we used BLEU, which compares the degree of overlap between n-grams in generated translations and one or more high-quality human reference translations \cite{papineni2002bleu}. Specifically, we set $n$ of n-grams to 4 and used $S_{\text{BLEU-1}}$, $S_{\text{BLEU-2}}$, $S_{\text{BLEU-3}}$ and $S_{\text{BLEU-4}}$ to evaluate the generated outputs carefully and comprehensively.

\textbf{ROUGE}. ROUGE is mainly used to evaluate the quality of automatic text summarization and machine translation \cite{lin2004rouge}. In this paper, we used both ROUGE-n and ROUGE-l to assess the extent to which the generated text covers key information in the standard answer. In terms of ROUGE-n, the score is calculated based on n-gram overlap between generated text and referenced text. Different from ROUGE-n, ROUGE-l compares the longest common subsequence (LCS) of the prediction and the reference. We used $S_{\text{ROUGE-1}}$, $S_{\text{ROUGE-2}}$, and $S_{\text{ROUGE-l}}$ to represent the score of ROUGE-1, ROUGE-2 and ROUGE-l respectively.
    
\textbf{CIDEr}. We used CIDEr to calculate the cosine similarity between the predicted text and the Term Frequency-Inverse Document Frequency (TF-IDF) vectors of reference answers in the entire dataset to evaluate the consensus of the text \cite{vedantam2015cider}. In this paper, the CIDEr score is represented with $S_{\text{CIDEr}}$ .

\subsubsection{Hierarchical and grouping analysis} 

To move beyond a single, aggregated performance metric, our evaluation framework incorporates a hierarchical and grouping analysis, which enables a more granular assessment of the capabilities of the model. The process begins by calculating the single value of the above metrics for each piece of SQA data, so each piece of data under each signal group (named after the signal type) will receive a score. Then the scores for each metric of each piece of data under each signal group are summed and averaged to obtain the average score for each metric of each signal group. Afterwards, for the entire evaluation dataset, the average scores of each signal group under each metric are summed and averaged to obtain the overall score. Finally, all of the average values are summarized in the evaluation report. This decomposition allows for a detailed analysis of the performance on specific tasks, such as identifying the amplitude of a single harmonic signal. This evaluation method facilitates the pinpointing of the specific strengths and weaknesses of the model, revealing whether VSLLaVA performs better on harmonic signals compared to modulated signals, or whether VSLLaVA has a general difficulty with parameter identification tasks.

The complete evaluation framework is detailed in Algorithm \ref{alg:main_evaluation}. It takes the SQA evaluation dataset, denoted as $\mathcal{D}$, and the referee model configuration, $C$, as inputs. Each sample $d \in \mathcal{D}$ consists of a question $X_{q}$, a model-generated prediction $X_{p}$, and the corresponding standard answer $X_{a}$. The evaluation proceeds in following steps. First, for the LLM-based assessment, an evaluation input is constructed for each sample by combining $X_{q}$, $X_{p}$, and $X_{a}$ with a predefined prompt template. This composite input is then processed by the external referee LLM, loaded according to configuration $C$, to produce a score $S_{\text{llm}}$. Second, for each piece of SQA data, the Numerical Score $S_n$ and the Word Recall $S_w$ are calculated based on Equations \ref{eq:relative_error} to \ref{eq:word_recall}. Third, the suite of semantic evaluation metrics, including $S_\text{BLEU-1}$, $S_\text{BLEU-2}$, $S_\text{BLEU-3}$, $S_\text{BLEU-4}$, $S_\text{ROUGE-1}$, $S_\text{ROUGE-2}$, $S_\text{ROUGE-l}$ and $S_\text{CIDEr}$, is computed for each piece of data to provide a multi-faceted view of linguistic similarity, fluency, and consensus. Finally, the model performance is evaluated by a hierarchical process, starting from calculating the average value of each metric for each SQA group of different signal types to calculating the average value of each metric for the entire dataset, with each group macro-averaged. 

This granular assessment allows for a detailed evaluation of model capabilities across different sub-tasks. All individual scores, overall averages, and hierarchical results are serialized into structured files to support robust error analysis and subsequent model iterations.

\begin{algorithm}[]
\caption{Dual-mode Model Evaluation (Main Process)}
\label{alg:main_evaluation}
\begin{algorithmic}[1]

\Function{HybridEvaluation}{the evaluation dataset $\mathcal{D}$, the referee model configuration file $C$} 
    \State \Comment{See Table \ref{tab:notations_comprehensive} for notations}
    \State Initialize referee LLM evaluation results $R_\text{llm}$
    \State Initialize custom metrics evaluation results $R_\text{custom}$
    \State Initialize specialized analysis results $R_\text{analysis}$
    \Statex
    
    \State $R_\text{llm} \gets$ \textsc{ScoreWithRefereeLLM}($\mathcal{D}, C$) \Comment{See Algorithm \ref{alg:referee_scoring}}
    \Statex
    
    \State $R_\text{custom} \gets$ \textsc{CalculateRuleBasedMetrics}($\mathcal{D}$) \Comment{See Algorithm \ref{alg:rule_based_metrics}}
    \Statex
    
    \State Group $R_\text{custom}$ by \textit{signal category} and \textit{question} \Comment{\textit{Hierarchical and Grouping Analysis}}
    \For{each group in grouped results}
        \State Calculate average scores $S_\text{avg\_group}$ of all metrics in each group
        \State Save $S_\text{avg\_group}$ to $R_\text{analysis}$
    \EndFor
    \State Calculate overall scores $S_\text{avg\_overall}$ for all metrics in $R_\text{custom}$    
    \Statex
    
    \State Construct final report $R_\text{final} \gets (R_\text{llm}, R_\text{custom}, R_\text{analysis})$ 
    \State \Return $R_{final}$
\EndFunction

\end{algorithmic}
\end{algorithm}

\begin{algorithm}[h]
\caption{Helper Function: ScoreWithRefereeLLM}
\label{alg:referee_scoring}
\begin{algorithmic}[1]

\Function{ScoreWithRefereeLLM}{the evaluation dataset $\mathcal{D}$, the referee model configuration file $C$} 
    \State \Comment{See Table \ref{tab:notations_comprehensive} for notations}
    \State Initialize $R_\text{llm}$
    \If{$C$ is not empty}
        \State \textbf{try}
        \State \quad Load the external referee model from $C$
        \State \quad Construct evaluation inputs from $\mathcal{D}$
        \State \quad Evaluate based on evaluation inputs
        \State \quad Calculate scores $S_\text{llm}$ 
        \State \quad Save $S_\text{llm}$ to $R_\text{llm}$
        \State \textbf{catch} Exception
        \State \quad Log errors to $R_\text{llm}$
        \State \textbf{end try}
    \Else
        \State Mark $R_\text{llm}$ as "skipped"
    \EndIf
    \State \Return $R_\text{llm}$
\EndFunction

\end{algorithmic}
\end{algorithm}

\begin{algorithm}[]
\caption{Helper Function: CalculateRuleBasedMetrics}
\label{alg:rule_based_metrics}
\begin{algorithmic}[1]

\Function{CalculateRuleBasedMetrics}{the evaluation dataset $\mathcal{D}$} \Comment{See Table \ref{tab:notations_comprehensive} for notations}
    \State Initialize custom metrics results $R_{custom}$
    \For{each sample $d = (X_{q}, X_{p}, X_{a})$ in $\mathcal{D}$}
        \State Calculate the Numerical Score $S_n$ \Comment{using Eq. \ref{eq:relative_error} to Eq. \ref{eq:number_score}}
        \State Calculate the Word Recall Score $S_w$ \Comment{using Eq. \ref{eq:word_recall}}
        \State Calculate the semantic scores $S_\text{BLEU-1}, S_\text{BLEU-2}, S_\text{BLEU-3}, S_\text{BLEU-4}$ 
        \State Calculate the semantic scores $S_\text{ROUGE-1}, S_\text{ROUGE-2}, S_\text{ROUGE-l}$ 
        \State Establish result record $R$ containing metadata and all calculated scores for sample $d$
        \State Add $R$ to $R_\text{custom}$
    \EndFor
    \Statex
    \State \Return $R_\text{custom}$
\EndFunction

\end{algorithmic}
\end{algorithm}

\section{Experiments}               
\label{experiments}

\subsection{Experiment Setup}     

To ensure ease of implementation and reproducibility, the training and evaluation were conducted based on ms-Swift \cite{zhao2024swiftascalablelightweightinfrastructure} and Evalscope \cite{evalscope_2024} framework respectively. Relevant parameters are indicated in Table \ref{training_parameters}. We divided SQA triples into a training set, a validation set, and a testing set to evaluate the performance of the model. We used Ovis2-8B as our baseline model \cite{lu2024ovis} and a different GLM-4.1V-Thinking-Flash \cite{hong2025glm} as the referee model for collaborative optimization with experts. Compared methods included InternVL3-1B, InternVL3-8B \cite{chen2024expanding,wang2024mpo,chen2024far,chen2024internvl} and llama3-llava-next-8b \cite{li2024llavanext-strong}. All experiments were conducted on our server equipped with 8 Nvidia 4090 GPUs.

\begin{table}[htbp ]
    \centering
    \caption{Hyperparameter settings for the SFT, Evaluation, and GRPO stages.}
    \label{training_parameters}
    \begin{tabular}{llccc}
        \toprule
        \textbf{Category} & \textbf{Parameter} & \textbf{SFT} & \textbf{Evaluation} & \textbf{GRPO} \\
        \midrule
        & Input Image Size & 336 & 336 & 336 \\
        Model Configuration & Max Sequence Length & 2048 & 2048 & --- \\
        & Max Completion Length & --- & 512 & 512 \\
        \midrule
        & Epochs & 5 & --- & 1 \\
        Training Hyperparameters & Learning Rate & 5e-5 & --- & 1e-6 \\
        & Warmup Ratio & 0.1 & --- & 0.01 \\
        & Weight Decay & 0.01 & --- & --- \\
        \midrule
        & LoRA Rank (r) & 8 & --- & --- \\
        LoRA Configuration & LoRA Alpha ($\alpha$) & 32 & --- & --- \\
        & LoRA Dropout & 0.1 & --- & --- \\
        \midrule
        & Temperature & --- & 0.0 & 0.5 \\
        Generation Parameters & Num Generations & --- & --- & 7 \\
        & GRPO Beta ($\beta$) & --- & --- & 0.1 \\
        \bottomrule
    \end{tabular}
\end{table}

\subsection{Evaluation results}       

\begin{figure}      
\centerline{\includegraphics[width=\linewidth]{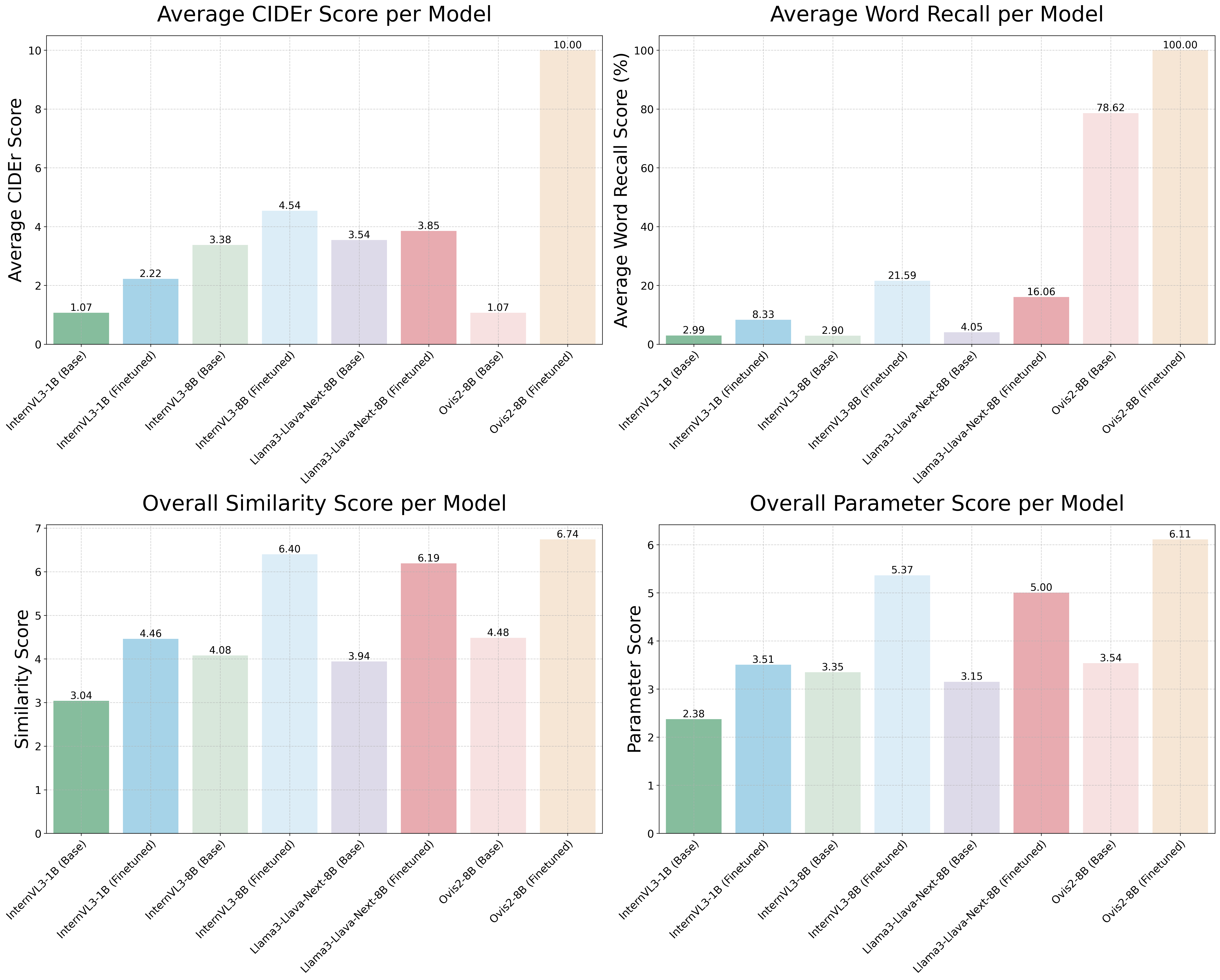}}
\caption{Scores of Average CIDEr and average Word Recall per model on the entire SQA dataset. In addition, we also utilized a referee model-GLM-4.1V-Thinking-Flash-to assess the models' abilities on our SQA dataset. The scores generated by the judging model ranged from \textbf{1} to \textbf{10}. The higher the score, the closer the generated answer is to the standard answer.}
\label{barplot}
\end{figure}

\begin{figure}      
\centerline{\includegraphics[width=\linewidth]{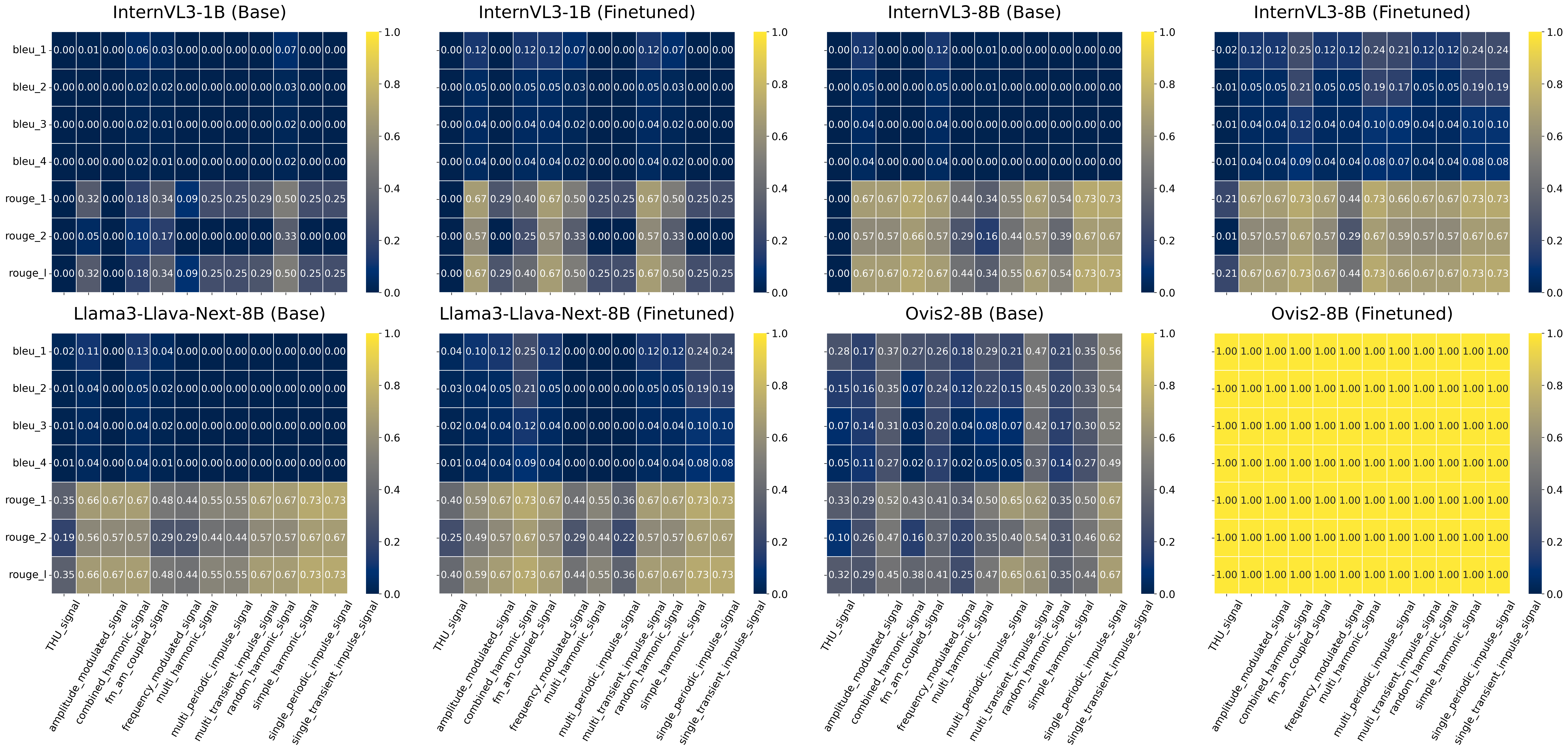}}
\caption{Scores of Average BLEU and average ROUGE per model on each SQA dataset. In each figure, the closer the color of the block is to yellow, the higher the score; the closer the color is to blue, the lower the score.}
\label{heatmap}
\end{figure}

Upon completion of the SFT phase, each model was systematically evaluated using the dual-mode framework detailed in Section \ref{evaluation_method}. The comprehensive results, including overall performance metrics and granular scores for each SQA category, are summarized in Table \ref{overall_score}, Fig. \ref{barplot}, and Fig. \ref{heatmap}.

\begin{table*}      
    \centering
    \caption{
        Performance comparison of different models at their Base and Fine-tuned stages. Our VSLLaVA utilizes Ovis2-8B as the baseline model.
        For Mean Relative Error, lower is better. 
        Note: llava-next-8B is the abbreviation for LLama3-llava-next-8B; Word Recall. is the abbreviation for Word Recall; Mean Rel. Err. is the abbreviation for Mean Relative Error; Num. Score is the abbreviation for Number Score. 
    }
    \label{overall_score}
    
    \setlength{\tabcolsep}{4pt} 
    
    \begin{tabular}{
        ll 
        S[table-format=2.2]
        S[table-format=4.2]
        S[table-format=1.2]
        S[table-format=1.2]
        c
        c
    }
        \toprule
        \multirow{2}{*}{\textbf{Model}} & \multirow{2}{*}{\textbf{Stage}} & \multicolumn{6}{c}{\textbf{Overall Score}} \\
        \cmidrule(l){3-8}
        & & {\shortstack{Word Rec. \\ (\%)}} & {\shortstack{Mean Rel. \\ Err.}} & {\shortstack{Num. \\ Score}} & {CIDEr} & {\shortstack{BLEU \\ (1/2/3/4)}} & {\shortstack{ROUGE \\ (1/2/L)}} \\
        \midrule
        
        \multirow{2}{*}{InternVL3-1B}
        & fine-tuned & 11.10 & 325.46  & 0.37 & 0.73 & 0.09 / 0.07 / 0.06 / 0.05 & 0.22 / 0.10 / 0.21 \\
        & base       & 7.24  & 17.28   & 0.33 & 0.31 & 0.06 / 0.04 / 0.03 / 0.02 & 0.14 / 0.05 / 0.13 \\
        \midrule
        
        \multirow{2}{*}{InternVL3-8B}
        & fine-tuned & 25.31 & 68.12   & 0.53 & 2.15 & 0.21 / 0.18 / 0.16 / 0.15 & 0.39 / 0.27 / 0.38 \\
        & base       & 9.22  & 159.79  & 0.36 & 0.69 & 0.07 / 0.04 / 0.03 / 0.03 & 0.21 / 0.11 / 0.20 \\
        \midrule
        
        \multirow{2}{*}{VSLLaVA}
        & fine-tuned &  \textbf{80.81} & 130.11 & \textbf{0.58} & \textbf{5.52} & \textbf{0.78 / 0.74 / 0.72 / 0.70} & \textbf{0.80 / 0.74 / 0.79} \\
        & base       & 64.29 & 981.82  & 0.31 & 0.16 & 0.20 / 0.14 / 0.10 / 0.08 & 0.34 / 0.18 / 0.30 \\
        \midrule
        
        \multirow{2}{*}{llava-next-8B}
        & fine-tuned & 16.80 & 1524.69 & 0.51 & 1.56 & 0.14 / 0.10 / 0.09 / 0.08 & 0.34 / 0.20 / 0.33 \\
        & base       & 16.00 & \bfseries 11.46   & 0.34 & 0.57 & 0.05 / 0.03 / 0.02 / 0.02 & 0.18 / 0.10 / 0.17 \\
        
        \bottomrule
    \end{tabular}
\end{table*}

Based on the results of overall performance shown in the Table \ref{overall_score}, it can be observed that except for the number relative error, other scores measuring the quality of the descriptive text of the four models all improved after fine-tuning, indicating that the fine-tuned the models experienced an enhancement in the performance of signal analysis. InternVL3-1B, InternVL3-8B, VSLLaVA, and LLama3-llava-next-8B experienced an increase of Word Recall by 3.86\%, 16.09\%, 16.52\%, and 0.8\%, respectively. This means that after fine-tuning, the model predictions are more consistent with the actual values. In terms of CIDEr scores, our VSLLaVA achieved the most significant improvement on this metric after fine-tuning, with a score of 5.52, indicating that the consensus between the generated subtitles and the reference subtitles is relatively higher.

Specifically, in terms of number relative error, this metric is used to measure the accuracy of the model in reading parameters in signal parameter identification tasks. It can be seen from Table \ref{overall_score} that both InternVL3-1B and Llama-llava-Next-8B experienced a dramatic rise in the number of relative error after fine-tuning, respectively 308.18 and 1513.23. For InternVL3-1B, this is relevant to the constraints caused by its small-scale model parameters, which hinder model performance on such complex tasks; for Llama-llava-Next-8B, the performance decline is probably on account of its language-vision connector. The original LLaVA-style models used a single linear layer as the connector. This structure may be too thin, making Llama-llava-Next-8B inevitable for the model to experience knowledge forgetting or knowledge confusion when learning complex tasks such as signal analysis, thus finally contributing to poor performance on parameter identification. While moderate-scale models like InternVL3-8B and VSLLaVA, they respectively adopted a two-layer MLP and an embedding table for modal alignment, thereby achieving better learning capabilities and better robustness. In addition, our VSLLaVA outperformed other models in terms of all semantic metrics after fine-tuning, and answered all questions about signal types correctly on our evaluation SQA dataset, as shown in Table \ref{overall_score} and Fig. \ref{barplot}, demonstrating potential in handling signal analysis tasks.

In Fig. \ref{heatmap}, the evaluation of BLEU and ROUGE scores on each SQA dataset is presented, reflecting the model performance on various signal types. It can be observed that all models experienced an improvement in signal analysis quality after fine-tuning. It is worth noting that, compared to the ROUGE score, the scores of each model on the four BLEU metrics are not particularly high. This is because the calculation of the BLEU score focuses more on precision, i.e., the proportion of words in the response that appear in the true value, requiring a perfect textual match without considering semantic and grammatical correctness. In contrast, the calculation of the ROUGE score emphasizes recall, primarily measuring whether the model output covers the key information in the true value. Precision and recall are usually mutually exclusive. Therefore, if the model output is semantically identical to the true value but differs in wording, the output will receive a very low BLEU score.

In addition, it is noticeable that in Fig. \ref{heatmap}, the ROUGE scores on the THU signal dataset is relatively lower than others. This may be because the THU dataset is the only real dataset among all datasets, which has no discernible pattern and contains a certain amount of noise. Without additional processing, it is difficult for the model to directly determine the signal type based on a simple time-domain waveform diagram and inadequate relevant knowledge of this signal type. Therefore, improving the ability of the model to identify real signals remains a challenge that requires the relentless efforts of researchers.

Furthermore, to reduce the subjectivity of expert rule-based evaluation methods, we also employed a referee model-GLM-4.1V-Thinking-Flash-to assess both standard and predicted responses. In this paper, the referee model assessed the generated results based on the similarity between the model generation and the standard answers, and the accuracy of parameter identification, which was evaluated from four perspectives: helpfulness, relevance, accuracy, and expertise. Scores ranged from 1 to 10, with higher scores indicating that the judging model considers the generated answers to be closer to the standards.

The results, as shown in Fig. \ref{barplot} and Table \ref{tab:model_scores}, indicated that the models fine-tuned on the SQA dataset achieve higher evaluation scores than the base models. Under the guidance of prompt words, the referee LLM acted as a professional vibration signal analyst, evaluating the predicted responses based on the requirements and the ground truth. 

\begin{table*}[htbp ]
    \centering
    \caption{The average score of the identification ability of each model on different SQA datasets presented by the referee model. Our VSLLaVA utilizes Ovis2-8B as the baseline model. Note that Sim. represents Similarity Score, Param. represents Parameter Score, llava-next-8B is the abbreviation for LLama3-llava-next-8B.}
    \label{tab:model_scores}
    \resizebox{\textwidth}{!}{
    \begin{tabular}{llcccccccccccc}
        \toprule
        \multicolumn{2}{c}{\multirow{2}{*}{\textbf{Model}}} & \multicolumn{2}{c}{\textbf{AM}} & \multicolumn{2}{c}{\textbf{FM}} & \multicolumn{2}{c}{\textbf{AMFM}} & \multicolumn{2}{c}{\textbf{SH}} & \multicolumn{2}{c}{\textbf{MH}} & \multicolumn{2}{c}{\textbf{RH}} \\
        \cmidrule(lr){3-4} \cmidrule(lr){5-6} \cmidrule(lr){7-8} \cmidrule(lr){9-10} \cmidrule(lr){11-12} \cmidrule(lr){13-14}
         & & Sim. & Param. & Sim. & Param. & Sim. & Param. & Sim. & Param. & Sim. & Param. & Sim. & Param. \\
        \midrule
        
        \multirow{2}{*}{InternVL3-1B} 
        & fine-tuned & 5.94 & 4.36 & 5.50 & 4.29 & 3.82 & 2.97 & 4.44 & 3.91 & 5.28 & 4.64 & 4.07 & 3.32 \\
        & base      & 3.40 & 2.51 & 3.07 & 2.46 & 2.66 & 1.99 & 3.77 & 3.30 & 2.36 & 1.91 & 2.53 & 2.02 \\
                                      
        \midrule 
        \multirow{2}{*}{InternVL3-8B} 
        & fine-tuned & \textbf{7.06} & 5.89 & 5.46 & 4.77 & 6.49 & 4.85 & 6.97 & 5.73 & \textbf{7.41} & 6.45 & 5.16 & 4.23 \\
        & base      & 3.35 & 2.89 & 3.20 & 2.78 & 3.74 & 3.05 & 3.86 & 3.20 & 5.03 & 4.02 & 4.16 & 3.34 \\
                                      
        \midrule
        \multirow{2}{*}{our VSLLaVA}     
        & fine-tuned & 6.59 & \textbf{6.25} & \textbf{6.61} & \textbf{6.17} & \textbf{7.60} & \textbf{7.19} & \textbf{8.61} & \textbf{7.00} & 7.06 & \textbf{6.81} & \textbf{6.59} & \textbf{5.77} \\
        & base      & 4.34 & 3.35 & 4.20 & 3.19 & 4.53 & 3.44 & 4.12 & 3.35 & 4.76 & 3.71 & 4.19 & 3.36 \\
                                      
        \midrule
        \multirow{2}{*}{llava-next-8B} 
        & fine-tuned& 5.48 & 4.47 & 5.89 & 4.31 & 5.18 & 4.29 & 8.28 & 5.92 & 8.47 & 6.89 & 5.07 & 4.26 \\
        & base     & 4.04 & 3.14 & 3.69 & 2.97 & 3.61 & 2.80 & 3.38 & 2.97 & 4.67 & 3.71 & 3.64 & 2.91 \\
                                       
        \midrule
        \midrule
        
        \multicolumn{2}{c}{\multirow{2}{*}{\textbf{Model}}} & \multicolumn{2}{c}{\textbf{CH}} & \multicolumn{2}{c}{\textbf{ST}} & \multicolumn{2}{c}{\textbf{MT}} & \multicolumn{2}{c}{\textbf{SP}} & \multicolumn{2}{c}{\textbf{MP}} & \multicolumn{2}{c}{\textbf{THU}} \\
        \cmidrule(lr){3-4} \cmidrule(lr){5-6} \cmidrule(lr){7-8} \cmidrule(lr){9-10} \cmidrule(lr){11-12} \cmidrule(lr){13-14}
         & & Sim. & Param. & Sim. & Param. & Sim. & Param. & Sim. & Param. & Sim. & Param. & Sim. & Param. \\
        \midrule

        \multirow{2}{*}{InternVL3-1B} 
        & fine-tuned & 4.81 & 3.29 & 4.83 & 3.60 & 2.79 & 2.12 & 4.88 & 3.93 & 3.44 & 2.36 & 3.62 & 3.12 \\
        & base      & 2.62 & 1.91 & 3.39 & 2.80 & 2.41 & 1.60 & 3.79 & 3.05 & 3.14 & 2.24 & 3.62 & 2.98 \\
                                      
        \midrule
        \multirow{2}{*}{InternVL3-8B} 
        & fine-tuned & 6.01 & 4.36 & \textbf{7.11} & \textbf{6.55} & 4.85 & 3.78 & 7.84 & \textbf{7.66} & \textbf{7.21} & 6.09 & \textbf{5.17} & \textbf{4.02} \\
        & base      & 5.47 & 4.27 & 4.07 & 3.85 & 4.16 & 2.86 & 3.77 & 3.49 & 3.87 & 2.94 & 4.13 & 3.51 \\
                                      
        \midrule
        \multirow{2}{*}{our VSLLaVA}  
        & fine-tuned & \textbf{6.67} & \textbf{5.76} & 6.16 & 5.58 & \textbf{4.96} & \textbf{4.36} & 7.78 & 7.43 & 7.11 & \textbf{6.73} & 4.75 & 3.85 \\
        & base      & 5.76 & 4.39 & 4.63 & 3.95 & 4.75 & 3.58 & 3.91 & 3.42 & 4.15 & 3.10 & 4.47 & 3.72 \\
                                      
        \midrule
        \multirow{2}{*}{llava-next-8B} 
        & fine-tuned & 6.17 & 4.91 & 5.40 & 4.80 & 4.69 & 3.58 & \textbf{7.89} & 6.94 & 6.64 & 5.63 & 4.63 & 3.67 \\
        & base     & 4.89 & 3.59 & 3.86 & 3.49 & 4.04 & 2.99 & 3.67 & 3.32 & 3.96 & 3.04 & 3.75 & 2.86 \\

        \bottomrule
    \end{tabular}
    } 
\end{table*}

\subsection{GRPO results}           

The initial fine-tuning phase enhances the performance of VSLLaVA on signal analysis tasks but also makes the model fall short of outputting concise and robust decisions required for an expert system. Consequently, our GRPO experiment served as a critical refinement stage. The primary goal was to sharpen the focus of VSLLaVA and steer its behavior away from the verbose, explanatory patterns learned during SFT and towards providing direct, accurate, and reliable signal type identifications.

Based on our SQA dataset, the ultimate objective was to enable the model to output the signal type corresponding to the image. Therefore, we modified the previously established SQA dataset by removing questions related to signal parameter identification and simplifying the SQA data, which only includes questions and answers about signal types. Using our customized reward function proposed in Section \ref{GRPO}, we set important parameters listed in Table \ref{training_parameters}. 

\begin{figure}[h!]      
\centerline{\includegraphics[width=\linewidth]{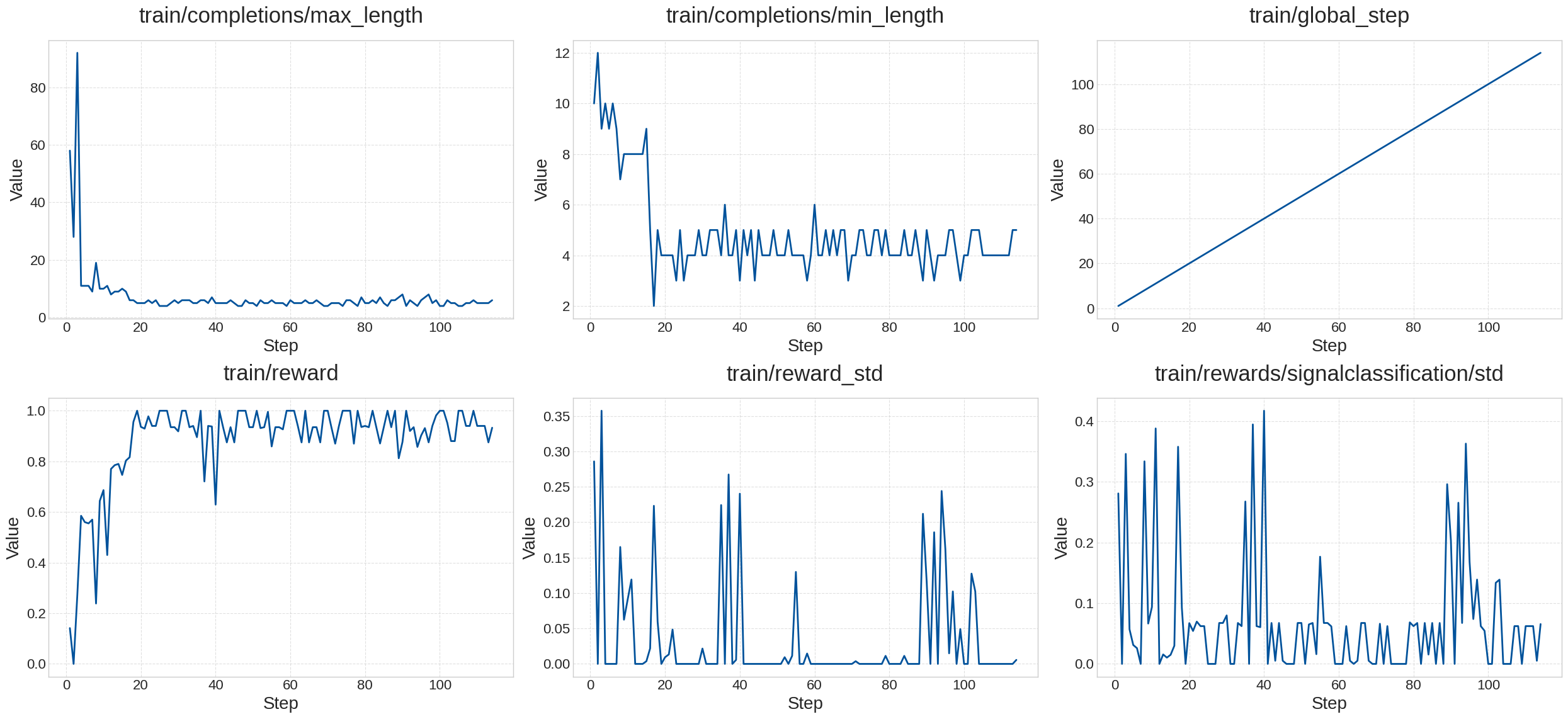}}
\caption{Changes in different indicators of VSLLaVA during GRPO training.}
\label{grpo_result}
\end{figure}

The experimental results provided compelling evidence that GRPO successfully reshaped the behavior of VSLLaVA, transforming the model from a "knowledgeable collaborator" into a "decisive expert." As shown in Fig. \ref{grpo_result}, our custom reward function increased as the training steps grew, proving that the model had learned how to complete this task, with the overall level fluctuating around 0.8 to 1.0. The standard deviation of reward eventually settled below 0.3 and has not yet fully converged, indicating that the model became more stable and reliable, yet still has room for improvement in terms of reward. 

Most significantly, the trend in completion length directly validated our primary motivation. The output length dramatically decreased from the initial state, stabilizing at approximately 5 tokens. This provided concrete proof that GRPO successfully mitigated the SFT-induced verbosity, forcing the model to produce the direct, "to-the-point" identifications we aimed for.  Furthermore, the qualitative examples in Table \ref{tab:grpo_case_study} revealed a deeper transformation. The model output exhibited an incremental reasoning process—akin to a human expert methodically analyzing an unfamiliar signal. This transition from simple pattern mimicry (SFT stage) to a structured, analytical approach (GRPO stage) confirmed a significant enhancement in the signal identification accuracy and robustness of the model.

\begin{table}[htbp ]
    \centering
    \caption{Case study of the GRPO reward mechanism. Note that this table only showcases examples of model-generated completions, the ground truth, and the best-matched synonym identified by the custom reward function.}
    \label{tab:grpo_case_study}
    
    \renewcommand{\tabularxcolumn}[1]{>{\small}m{#1}}
    \begin{tabularx}{\textwidth}{@{} c X @{}}
        \toprule
        \textbf{Ground Truth} & \textbf{Model Completion} \\
        \midrule
        
        \multirow{3}{*}{\texttt{Multiple Periodic Impulse Harmonic Signal.}} & 
        \textit{Thought}: <think> This is a multiple periodic impulse signal. </think>\\
        & \textbf{Answer}:\texttt{<answer> Multiple periodic impulse signal </answer>} \\
        \cmidrule(l){2-2} 
        & \textit{Best Match}: 'multiple periodic impulse harmonic signal' \\
        \midrule
        
        \multirow{3}{*}{\texttt{Combined Harmonic Signal.}} & 
        \textit{Thought}: <think>This is a combined harmonic signal, which means it is composed of multiple sine waves with different frequencies. Each vertical line represents the amplitude of a specific harmonic at different time points. </think>\\
        & \textbf{Answer}: \texttt{<answer> Combined harmonic signal. </answer>} \\
        \cmidrule(l){2-2}
        & \textit{Best Match}: 'combined harmonic signal' \\
        \midrule
        
        \multirow{3}{*}{\texttt{Single Transient Impulse Harmonic Signal.}} & 
        \textit{Thought}: <think>This is a single transient impulse signal. It decays over time with each impulse occurring at different time intervals. </think>\\
        & \textbf{Answer}:\texttt{<answer>Single transient impulse signal. </answer>} \\
        \cmidrule(l){2-2}
        & \textit{Best Match}: 'single transient impulse harmonic signal' \\
        \midrule
        
        \multirow{3}{*}{\texttt{Amplitude Modulated Signal.}} & 
        \textit{Thought}: <think>This is an amplitude modulated signal. It varies periodically with the modulation depth, which is represented by the variation in its amplitude over time. </think>\\
        & \textbf{Answer}: \texttt{<answer>Amplitude modulated signal. </answer>} \\
        \cmidrule(l){2-2}
        & \textit{Best Match}: 'amplitude modulated signal' \\

        \bottomrule
    \end{tabularx}
\end{table}

\section{Conclusion}
\label{conclusion}

This paper presents VSLLaVA, a specialized pipeline designed to imbue Large Multimodal Models (LMMs) with the domain-specific expertise required for industrial vibration signal analysis. VSLLaVA leverages expert-guided Signal-Question-Answer (SQA) triplets for parameter-efficient fine-tuning via Low-Rank Adaptation (LoRA), which effectively integrates signal processing knowledge into the LMM. A subsequent refinement stage using Group Relative Policy Optimization (GRPO) further enhances the classification robustness and response conciseness. Experimental results confirm that VSLLaVA significantly improves performance in signal type identification and parameter analysis of fault signals, demonstrating its potential as a foundational model for specialized industrial applications.

Despite these promising outcomes, this work has several limitations that open avenues for future research. First, the training data is predominantly synthetic. Although the real-world THU dataset was included, the generalization capability of the model to a wider array of noisy, real-world industrial signals requires further validation. Second, our current implementation exclusively fine-tunes the language model, leaving the vision encoder frozen. An encoder pre-trained on natural images may not be optimal for extracting salient features from one-dimensional signal visualizations, potentially creating a bottleneck in modality alignment. Future work could explore co-tuning the vision encoder or developing a signal-specific encoder to improve feature representation and overall model efficacy.

\section{Acknowledgements }
This work was supported by the National Natural Science Foundation of China 52305115.

\printcredits

\bibliographystyle{elsarticle-num}

\bibliography{bib1}





\clearpage 

\appendix 

\section*{Appendix} 
\addcontentsline{toc}{section}{Appendix} 

\section{Signal visualization of SQA triplets generators}
\label{app:vis_data}

Table \ref{Signal visualization} shows the 12 signals we used. It is worth noting that the signals shown are only examples. In our experiments, we generated many different sets of signals using random numbers based on the signal parameters shown in Table \ref{signal generators}. Out of these, 200 sets of each signal type were utilized for fine-tuning, while 20 sets of each signal type were allocated for evaluation and GRPO experiments.

\begin{table}[htbp ]
\centering
\caption{Signal visualization.}
\label{Signal visualization}
\begin{tabular}{>{\centering\arraybackslash}cccc}
\hline
\textbf{Dataset name}& \textbf{Signal visualization} & \textbf{Dataset name}&\textbf{Signal visualization} \\ 
\hline
    AM & \includegraphics[width=5cm]{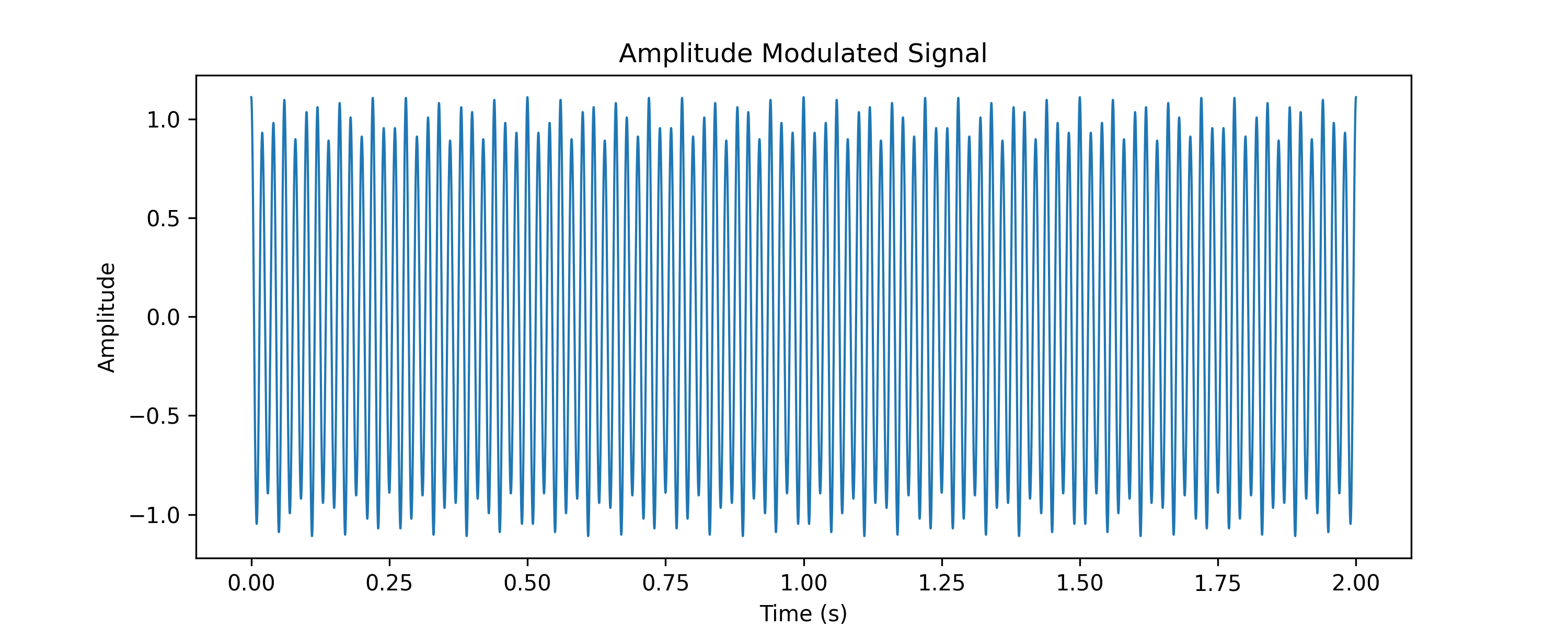} & 
    FM &\includegraphics[width=5cm]{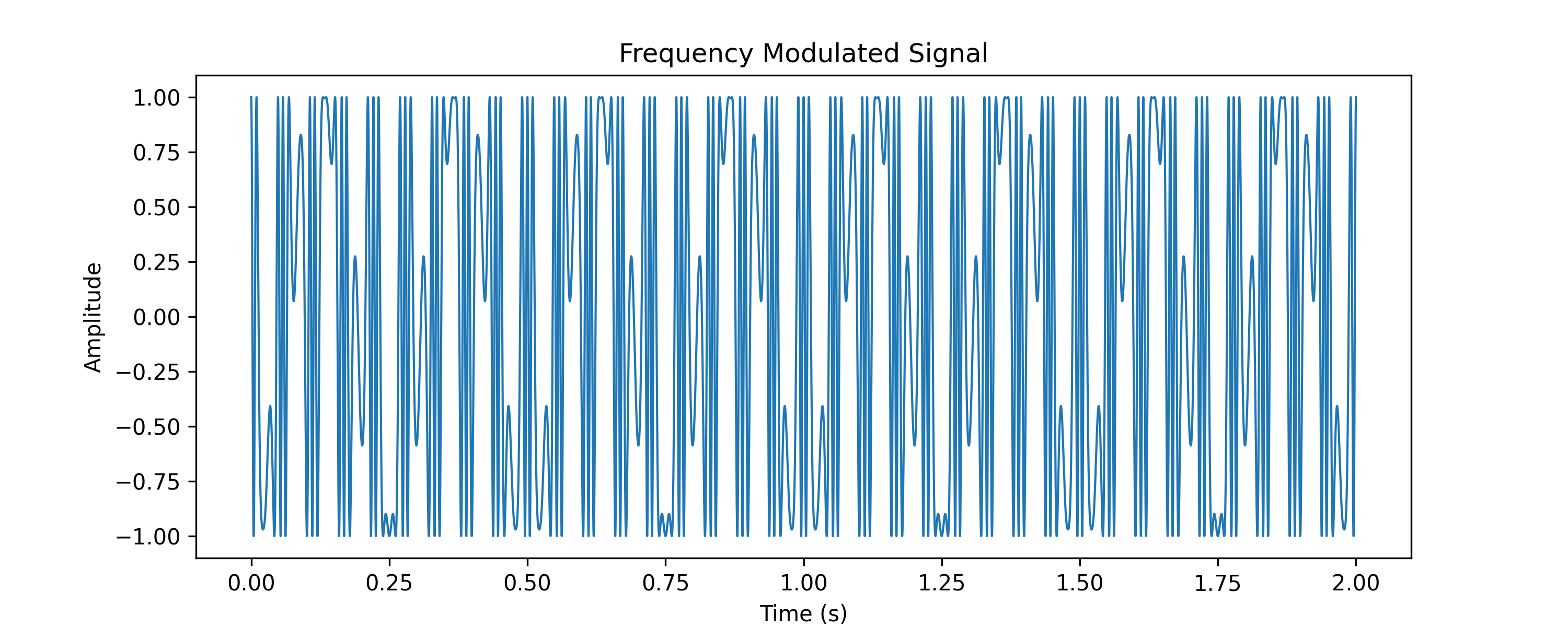}
\\
    AMFM & \includegraphics[width=5cm]{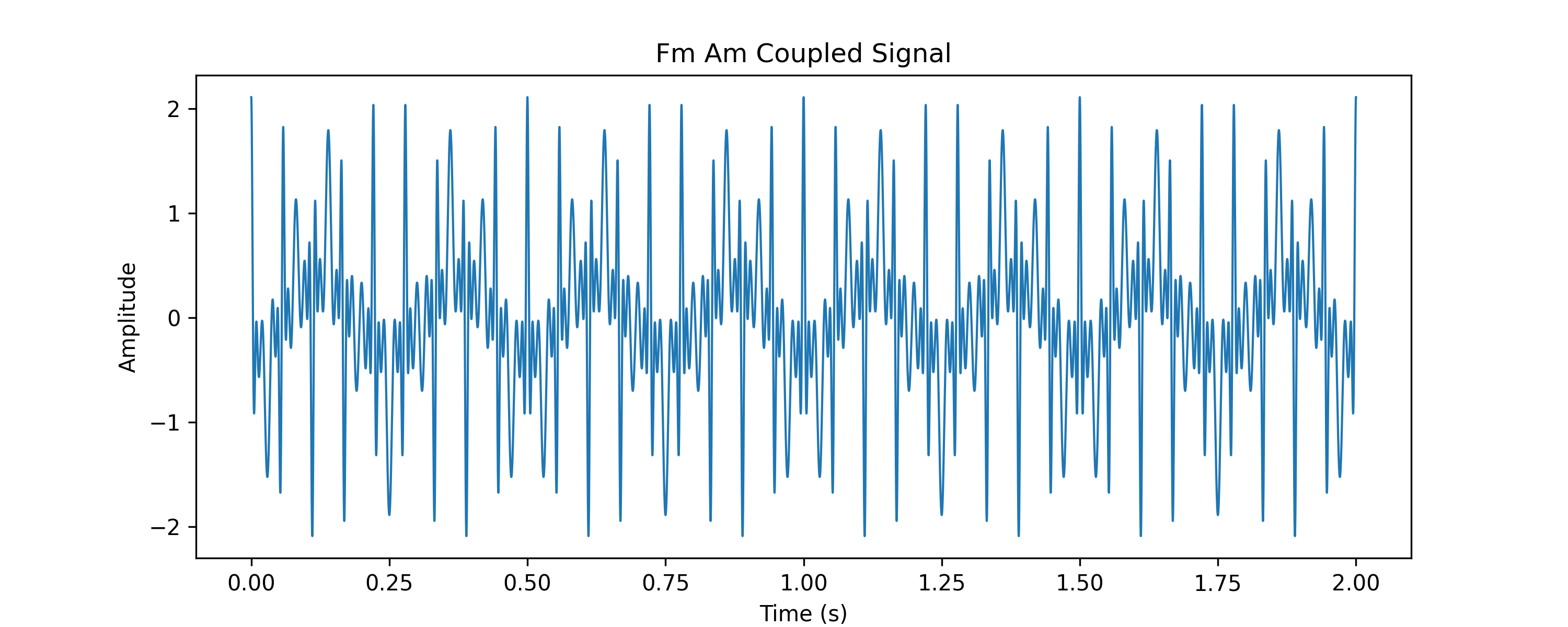} & 
    SH & \includegraphics[width=5cm]{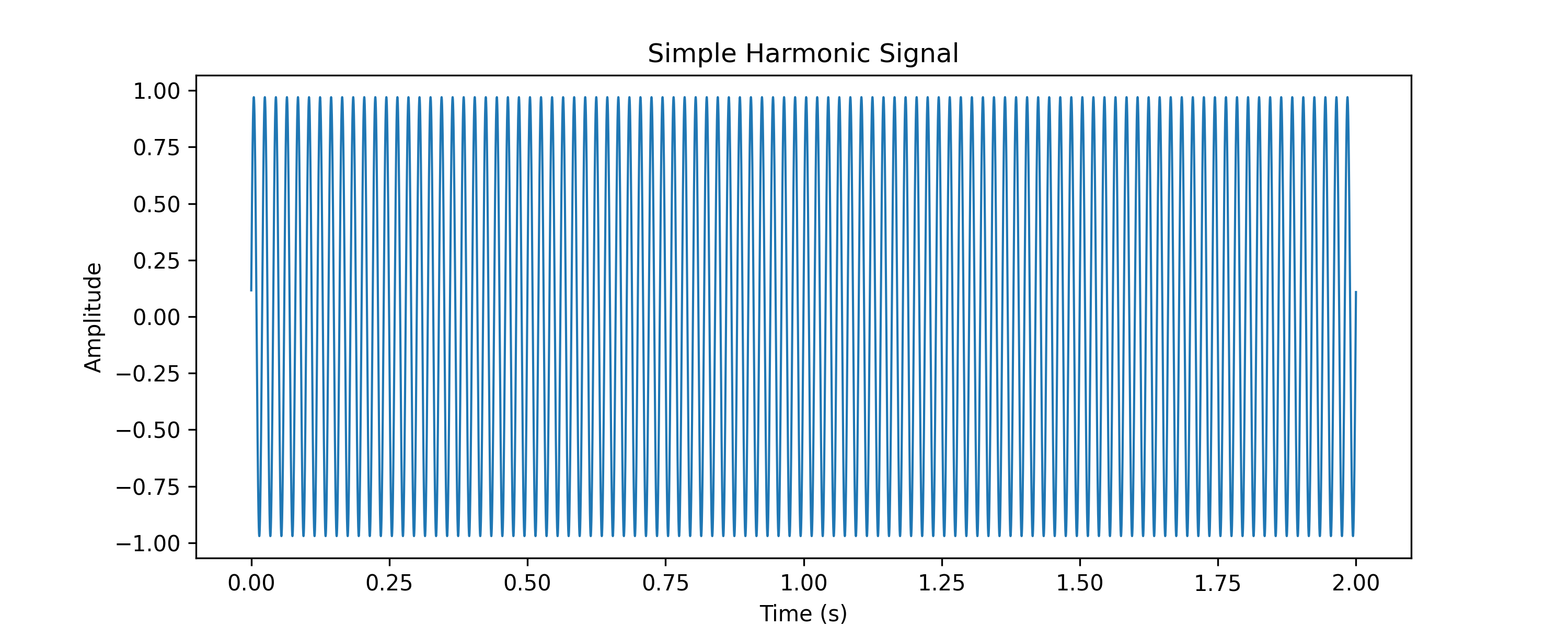}
\\
    MH &\includegraphics[width=5cm]{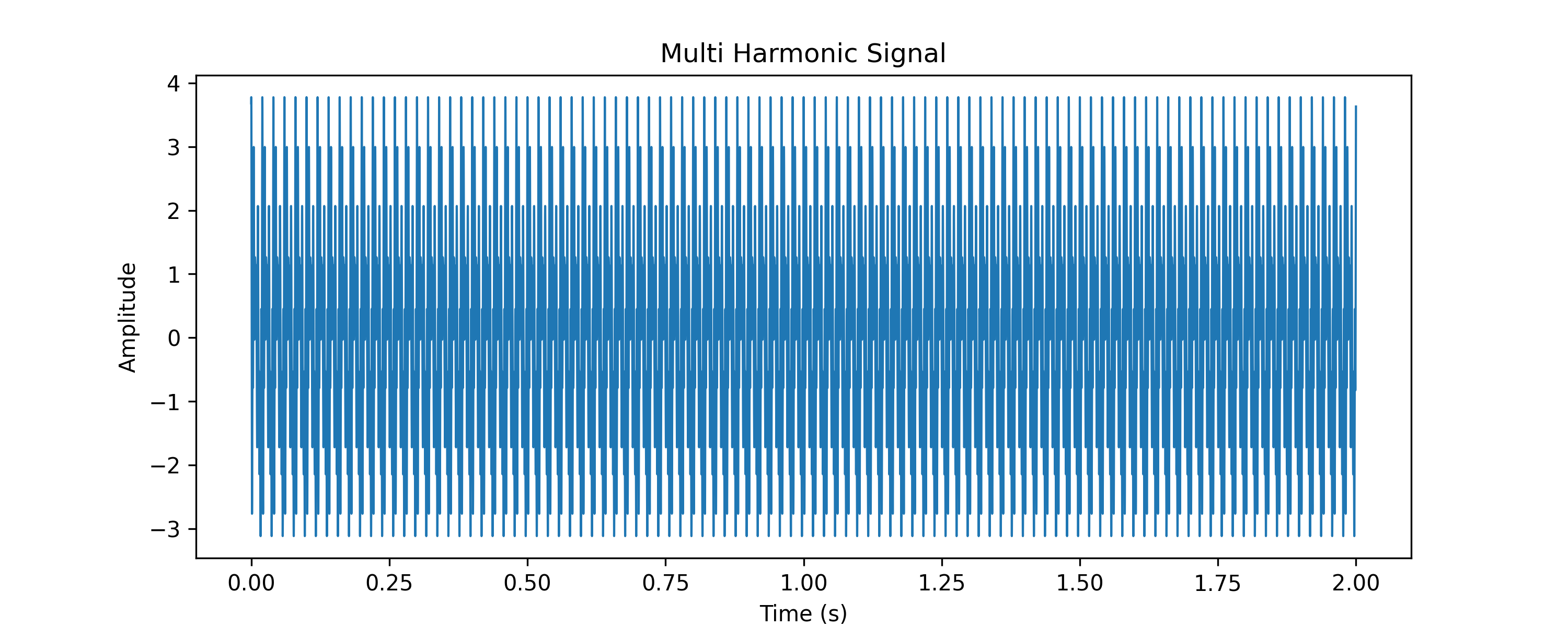} &
    RH &\includegraphics[width=5cm]{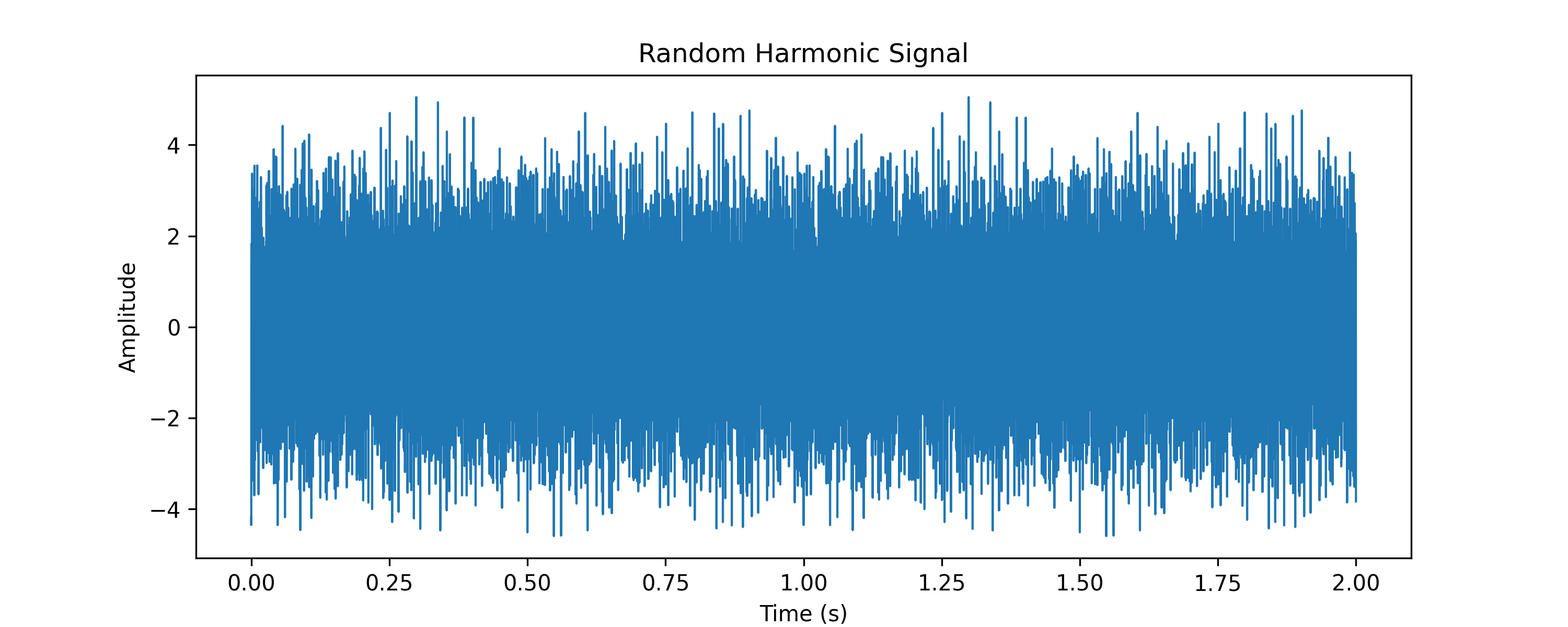} 
\\ 
    CH & \includegraphics[width=5cm]{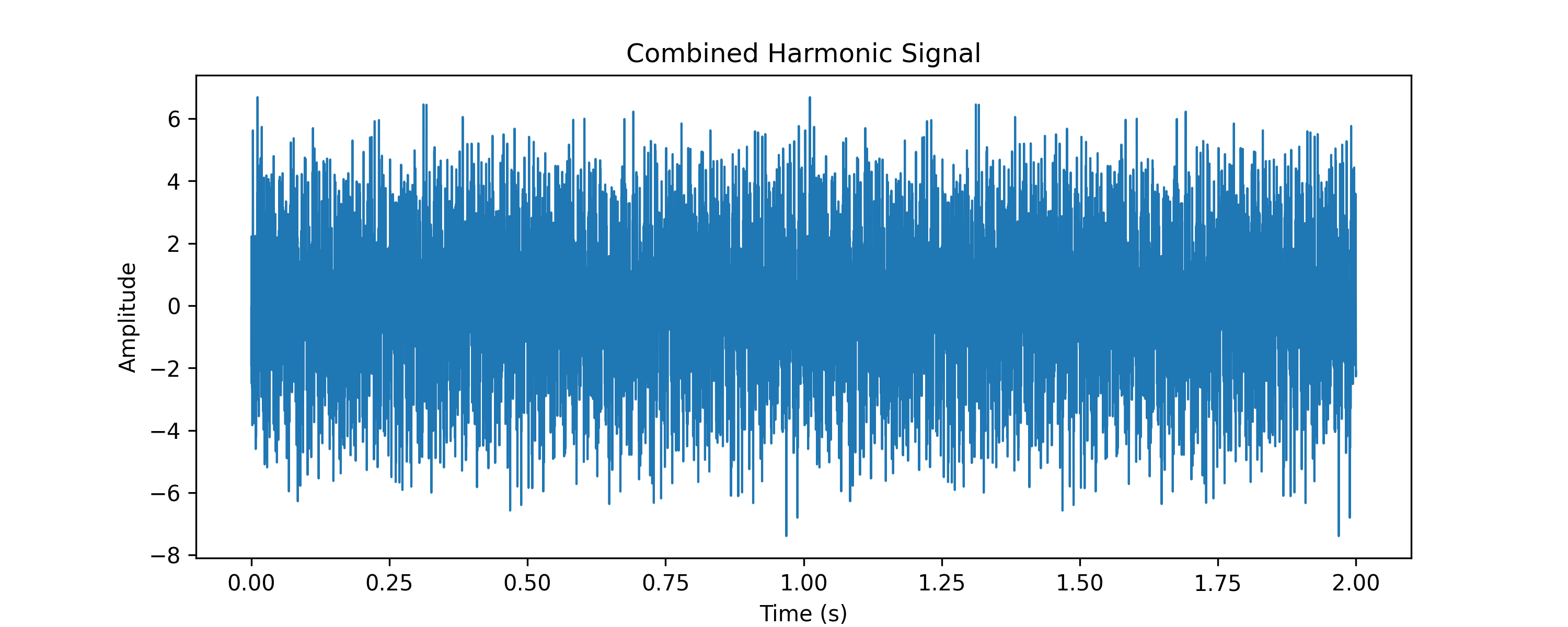} & 
    ST &\includegraphics[width=5cm]{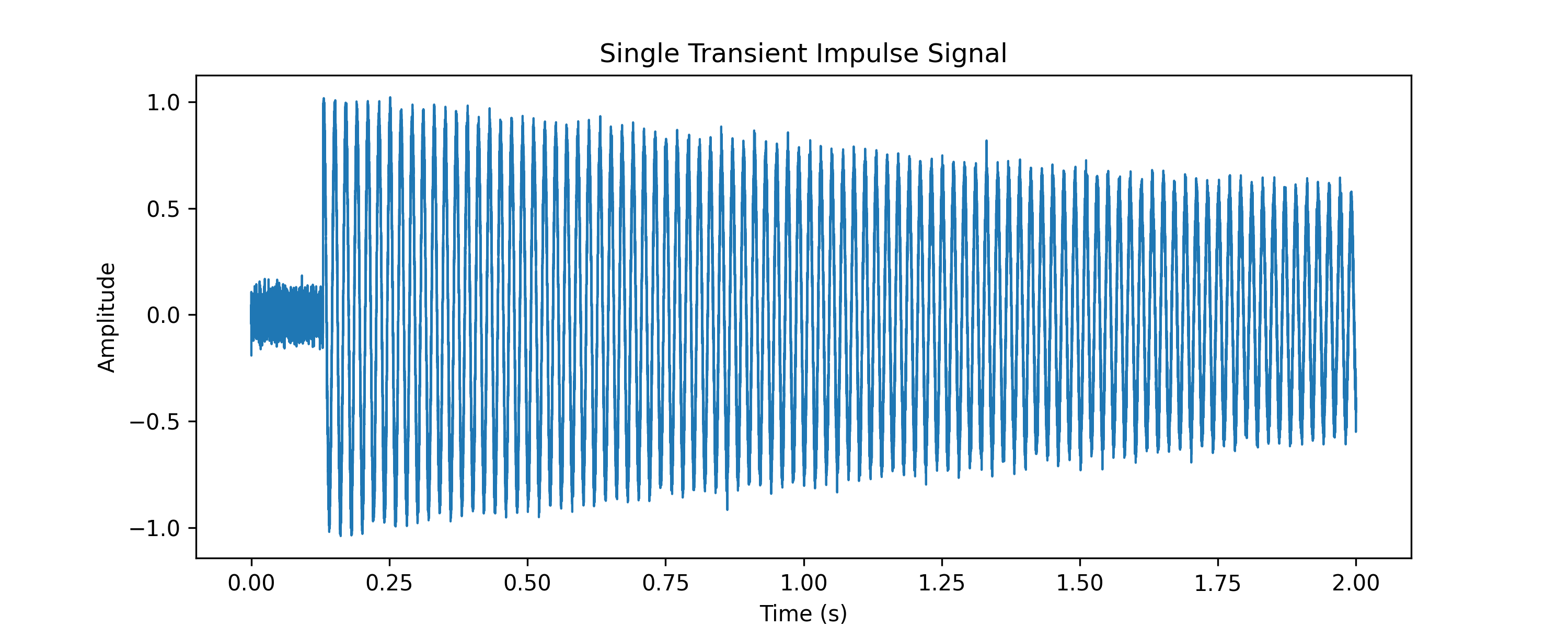}
\\
    MT &\includegraphics[width=5cm]{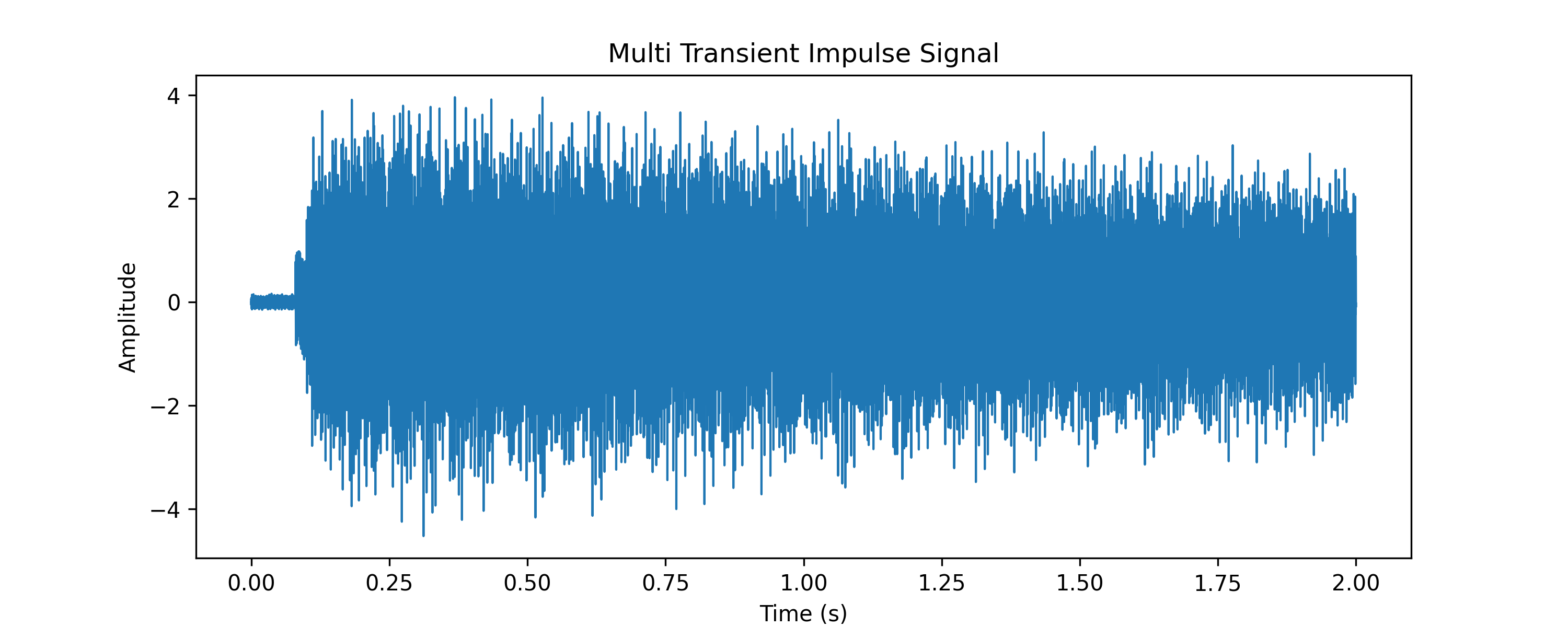} & 
    SP &  \includegraphics[width=5cm]{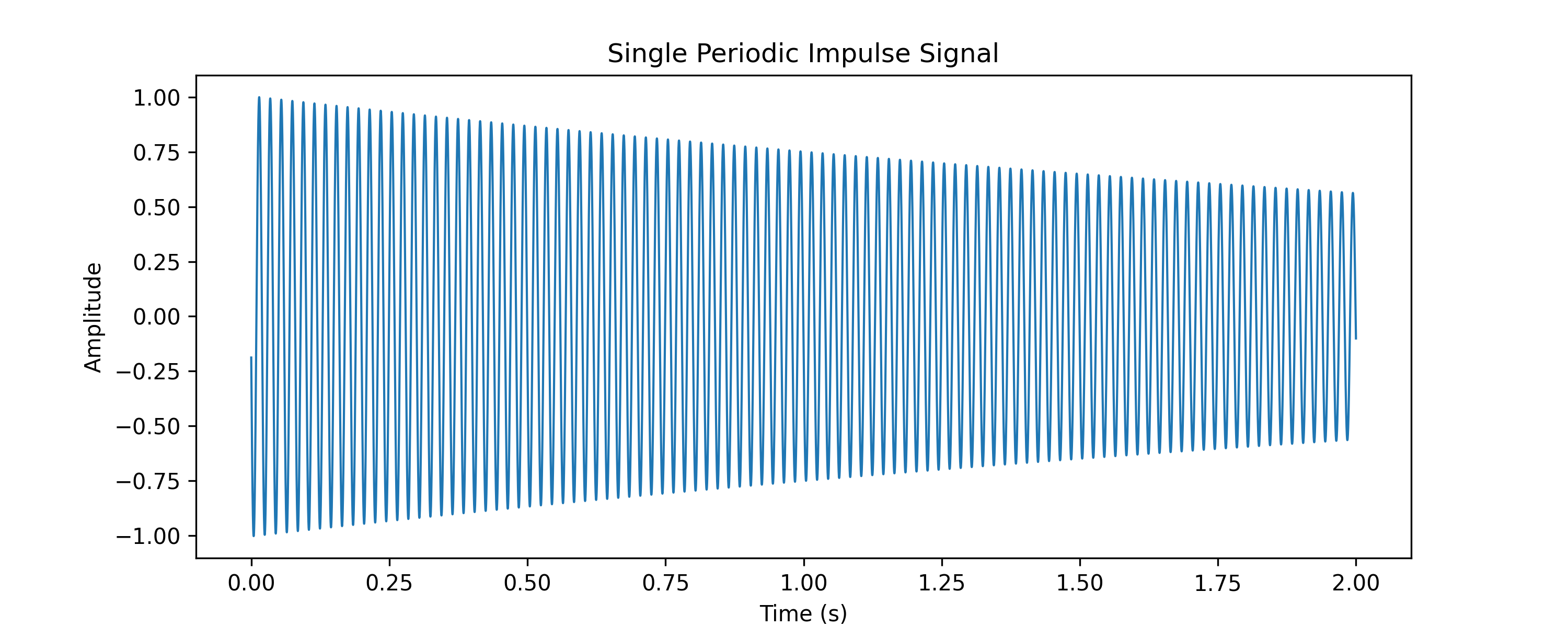} 
\\
    MP &\includegraphics[width=5cm]{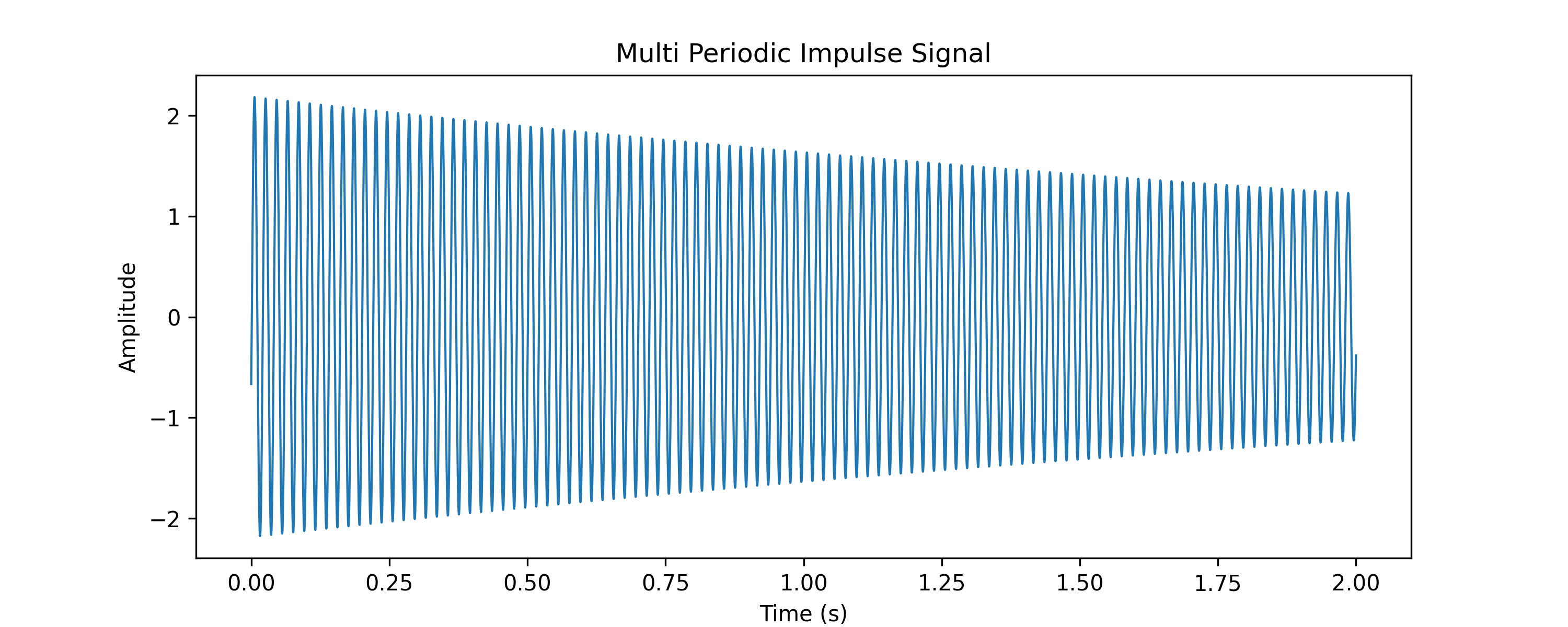} & THU &\includegraphics[width=5cm]{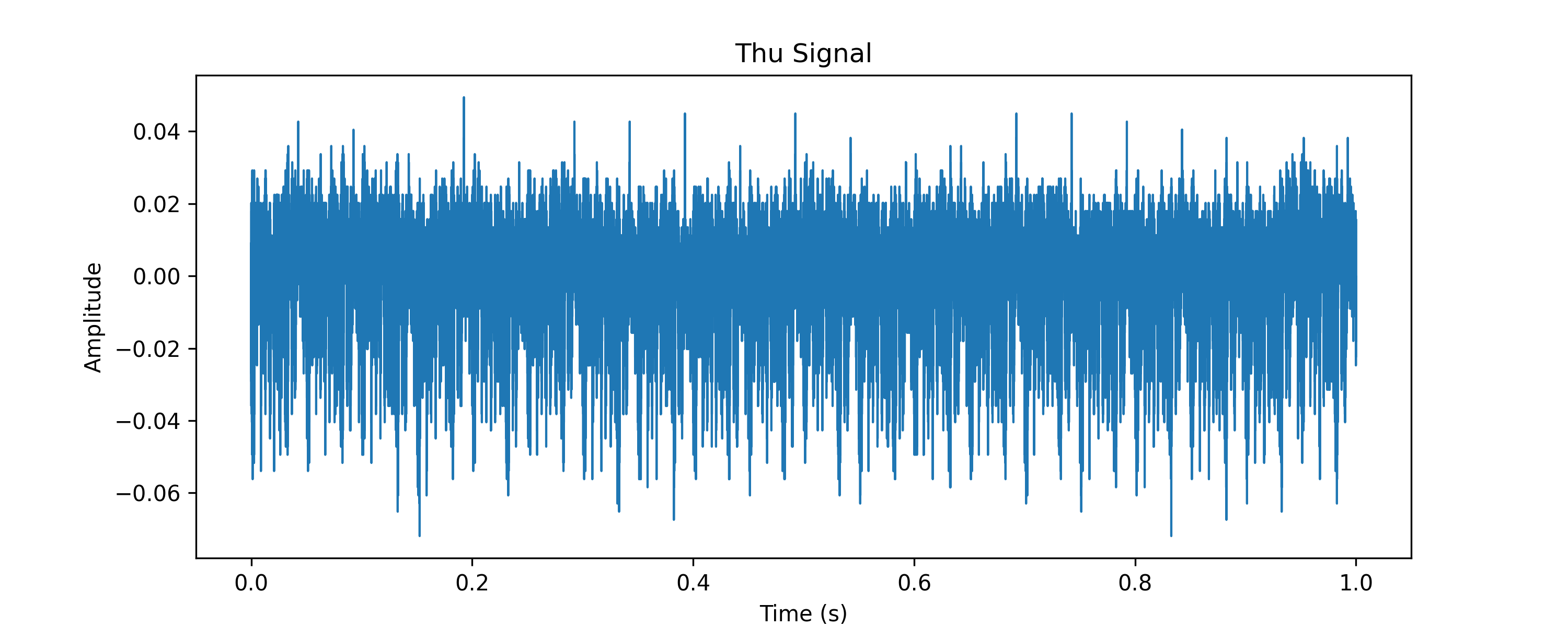}
\\ 

 \hline
\end{tabular}

\end{table}

\section{Synonym vocabulary and corresponding weights in GRPO experiments}
\label{app:GRPO_details}
Table \ref{tab:synonym_weights_part1} and Table \ref{tab:synonym_weights_part2} list the synonym vocabulary used in our GRPO experiments and the corresponding weights for each synonym. To generate these synonyms, we utilized Gemini 2.5 pro \cite{comanici2025gemini} to assist with brainstorming except for the THU signal. After obtaining the synonyms, we evaluated the quality and semantic relevance of the generated terms and assigned corresponding weights to each word. 

Here are the prompts for generating each signal:

"Please generate at least 10 terms with the same meaning as the given noun: Simple Harmonic Signal (single sinusoidal wave signal)."

"Please generate at least 10 terms with the same meaning as the given noun: Multiple Harmonic Signal (a signal composed of multiple superimposed harmonic sine waves)."

"Please generate at least 10 terms with the same meaning as the given term: Random Harmonic Signal (a signal composed of multiple superimposed sine waves with random fundamental frequencies)."

"Please generate at least 10 terms with equivalent meanings based on the given noun: Combined Harmonic Signal (a signal formed by adding the Multiple Harmonic Signal and Random Harmonic Signal mentioned above)."

"Please generate at least 10 terms with the same meaning as the given term: Frequency Modulated Signal."

"Please generate at least 10 terms with the same meaning as the given term: Amplitude Modulated Signal."

"Please generate at least 10 terms with equivalent meanings based on the given noun: FM-AM Coupled Signal (a signal formed by the superposition of FM and AM signals sharing the same carrier frequency)."

"Please generate at least 10 terms with equivalent meanings based on the given noun: Single Periodic Impulse Harmonic Signal (a single decaying signal with periodic impulse characteristics, as illustrated in the formula)."

"Please generate at least 10 terms with equivalent meanings based on the given noun: Multiple Periodic Impulse Harmonic Signal (a signal formed by the superposition of multiple decaying signals with periodic impulse characteristics, as illustrated in the formula)."

"Please generate at least 10 terms with equivalent meanings based on the given noun: Single Transient Impulse Harmonic Signal (a decaying signal exhibiting transient impulse characteristics, as illustrated in the formula)."

"Please generate at least 10 terms with equivalent meanings based on the provided noun: Multiple Transient Impulse Harmonic Signal (a signal composed of multiple superimposed decaying signals with transient impulse characteristics, as illustrated in the formula)."


\begin{table}[!Htbp]
\caption{Synonym Vocabulary and Corresponding Weights}
\label{tab:synonym_weights_part1}
\begin{tabular}{p{3.2cm} p{4.5cm} l p{4.5cm} l}
\toprule
\textbf{Signal Type} & \textbf{Synonym} & \textbf{Weight} & \textbf{Synonym} & \textbf{Weight} \\
\midrule
\multirow{5}{*}{\parbox{3.2cm}{Simple Harmonic Signal}} & Simple Harmonic Signal & 1.0 & Simple Harmonic & 1.0 \\
 & Single Harmonic Signal & 1.0 & Single Harmonic Signal & 1.0 \\
 & Simple Harmonic Wave & 0.9 & Single Harmonic Wave & 0.9 \\
 & Sinusoidal Signal & 0.8 & Sinusoidal Wave & 0.8 \\
 & Cosine Wave & 0.5 & Cosinusoidal Signal & 0.5 \\
\midrule
\multirow{5}{*}{\parbox{3.2cm}{Random Harmonic Signal}} & Random Harmonic Signal & 1.0 & Random Harmonic & 1.0 \\
 & Random Harmonic Wave & 0.9 & Stochastic Harmonic Wave & 0.9 \\
 & Stochastic Harmonic Signal & 0.8 & Stochastic Harmonic & 0.8 \\
 & Random Sinusoidal Signal & 0.5 & Stochastic Sinusoidal Signal & 0.4 \\
 & Random Sinusoidal Wave & 0.3 & Stochastic Sinusoidal Wave & 0.3 \\
\midrule
\multirow{5}{*}{\parbox{3.2cm}{Frequency Modulated Signal}} & Frequency Modulated Signal & 1.0 & FM Signal & 1.0 \\
 & Frequency Modulation Signal & 0.8 & Signal with Variable Instantaneous Frequency & 0.8 \\
 & Signal with Frequency Variation & 0.8 & Signal with Frequency Modulation & 0.8 \\
 & Angle-Modulated Signal & 0.7 & Angle Modulated Signal & 0.7 \\
 & Constant Envelope Signal & 0.4 & Constant Amplitude Signal & 0.4 \\
\midrule
\multirow{5}{*}{\parbox{3.2cm}{FM-AM Coupled Signal}} & FM-AM Coupled Signal & 1.0 & AM-FM Coupled Signal & 1.0 \\
 & Coupled FM-AM Signal & 1.0 & Coupled AM-FM Signal & 1.0 \\
 & Hybrid FM-AM Signal & 0.9 & Hybrid AM-FM Signal & 0.9 \\
 & Combined FM-AM Signal & 0.8 & Combined AM-FM Signal & 0.8 \\
 & Signal with Simultaneous Amplitude and Frequency Modulation & 0.7 & Complex Modulated Signal & 0.5 \\
\midrule
\multirow{5}{*}{\parbox{3.2cm}{Multiple Periodic Impulse Harmonic Signal}} & Multiple Periodic Impulse Harmonic Signal & 1.0 & Multiple Periodic Impulse Harmonic & 1.0 \\
 & Multiple Periodic Impulse Harmonic Wave & 0.9 & Multiple Periodic Impulse Harmonic Wave & 0.9 \\
 & Multiple Periodic Impulse Harmonic Oscillation & 0.8 & Multiple Periodic Impulse Harmonic Oscillation & 0.8 \\
 & Multiple Periodic Impulse Harmonic Response & 0.7 & Multiple Periodic Impulse Harmonic Response & 0.7 \\
 & Multiple Periodic Impulse Harmonic Signal with Damping & 0.5 & Multiple Periodic Impulse Harmonic Signal with Decay & 0.5 \\
\midrule
\multirow{5}{*}{\parbox{3.2cm}{Multiple Transient Impulse Harmonic Signal}} & Multiple Transient Impulse Harmonic Signal & 1.0 & Multiple Transient Impulse Harmonic & 1.0 \\
 & Multiple Transient Impulse Harmonic Wave & 0.9 & Multiple Transient Impulse Harmonic Wave & 0.9 \\
 & Multiple Transient Impulse Harmonic Oscillation & 0.8 & Multiple Transient Impulse Harmonic Oscillation & 0.8 \\
 & Multiple Transient Impulse Harmonic Response & 0.7 & Multiple Transient Impulse Harmonic Response & 0.7 \\
 & Multiple Transient Impulse Harmonic Signal with Damping & 0.5 & Multiple Transient Impulse Harmonic Signal with Decay & 0.5 \\
\bottomrule
\end{tabular}
\end{table}

\begin{table}[!Htbp]
\caption{Synonym Vocabulary and Corresponding Weights (continued from Table \ref{tab:synonym_weights_part1}).}
\label{tab:synonym_weights_part2}
\begin{tabular}{p{3.2cm} p{4.5cm} l p{4.5cm} l}
\toprule
\textbf{Signal Type} & \textbf{Synonym} & \textbf{Weight} & \textbf{Synonym} & \textbf{Weight} \\
\midrule
\multirow{5}{*}{\parbox{3.2cm}{Multiple Harmonic Signal}} & Multiple Harmonic Signal & 1.0 & Multiple Harmonic & 1.0 \\
 & Multi-Harmonic Signal & 1.0 & Multi-Harmonic & 1.0 \\
 & Multiple Harmonic Wave & 0.9 & Multi-Harmonic Wave & 0.9 \\
 & Complex Periodic Signal & 0.3 & Complex Periodic Wave & 0.3 \\
 & Composite Wave & 0.3 & Composite Signal & 0.3 \\
\midrule
\multirow{5}{*}{\parbox{3.2cm}{Combined Harmonic Signal}} & Combined Harmonic Signal & 1.0 & Combined Harmonic & 1.0 \\
 & Hybrid Harmonic Signal & 1.0 & Hybrid Harmonic & 1.0 \\
 & Harmonic Signal with Randomness & 0.7 & Harmonic Signal with Stochasticity & 0.7 \\
 & Harmonic Signal with Noise & 0.3 & Harmonic Signal with Variability & 0.3 \\
 & Harmonic Signal with Random Components & 0.3 & Harmonic Signal with Randomness & 0.3 \\
\midrule
\multirow{5}{*}{\parbox{3.2cm}{Amplitude Modulated Signal}} & Amplitude Modulated Signal & 1.0 & AM Signal & 1.0 \\
 & Amplitude Modulation Signal & 1.0 & Signal with Variable Amplitude & 0.8 \\
 & Signal with Amplitude Variation & 0.8 & Signal with Amplitude Modulation & 0.8 \\
 & Envelope Modulated Signal & 0.6 & Envelope Modulation Signal & 0.6 \\
 & Constant Frequency Signal & 0.4 & Constant Frequency Modulation Signal & 0.4 \\
\midrule
\multirow{5}{*}{\parbox{3.2cm}{Single Periodic Impulse Harmonic Signal}} & Single Periodic Impulse Harmonic Signal & 1.0 & Single Periodic Impulse Harmonic & 1.0 \\
 & Single Periodic Impulse Harmonic Wave & 0.9 & Single Periodic Impulse Harmonic Wave & 0.9 \\
 & Single Periodic Impulse Harmonic Oscillation & 0.8 & Single Periodic Impulse Harmonic Oscillation & 0.8 \\
 & Single Periodic Impulse Harmonic Response & 0.7 & Single Periodic Impulse Harmonic Response & 0.7 \\
 & Single Periodic Impulse Harmonic Signal with Damping & 0.5 & Single Periodic Impulse Harmonic Signal with Decay & 0.5 \\
\midrule
\multirow{5}{*}{\parbox{3.2cm}{Single Transient Impulse Harmonic Signal}} & Single Transient Impulse Harmonic Signal & 1.0 & Single Transient Impulse Harmonic & 1.0 \\
 & Single Transient Impulse Harmonic Wave & 0.9 & Single Transient Impulse Harmonic Wave & 0.9 \\
 & Single Transient Impulse Harmonic Oscillation & 0.8 & Single Transient Impulse Harmonic Oscillation & 0.8 \\
 & Single Transient Impulse Harmonic Response & 0.7 & Single Transient Impulse Harmonic Response & 0.7 \\
 & Single Transient Impulse Harmonic Signal with Damping & 0.5 & Single Transient Impulse Harmonic Signal with Decay & 0.5 \\
\midrule
\multirow{6}{*}{\parbox{3.2cm}{THU Signal}} & THU Signal & 1.0 & THU bearing signal & 1.0 \\
 & THU data & 1.0 & THU bearing data & 1.0 \\
 & THU health bearing & 1.0 & THU inner fault & 1.0 \\
 & THU outer fault & 1.0 & THU roller fault & 1.0 \\
 & THU bearing health & 1.0 & THU bearing inner fault & 1.0 \\
 & THU bearing outer fault & 1.0 & THU bearing roller fault & 1.0 \\
\bottomrule
\end{tabular}
\end{table}

\section{Response examples of VSLLaVA on different SQA datasets} 

We created 5 to 9 SQA triplets for each signal type, covering questions about signal types, signal parameters, and summaries of signal characteristics. Due to space constraints, in this section we have selected the output results of VSLLaVA for only two representative signals—simple harmonic signal and THU signal—from the evaluation process as examples, as former has a simple composition and can intuitively reflect VSLLaVA's performance in signal recognition and parameter identification tasks and the latter can demonstrate VSLLaVA's ability to identify fault signals. The results are shown in Table \ref{tab:qualitative_comparison_part1} and \ref{tab:qualitative_comparison_part2}.

\begin{table*}[htbp ]
    \centering
    \caption{Examples of model predictions against standard answers for different signal analysis cases. Note that we only chose the results of three representative signals.}
    \label{tab:qualitative_comparison_part1}
    
    \small 
    
    \renewcommand{\arraystretch}{1.5}       
    
    \begin{tabularx}{\textwidth}{@{} >{\RaggedRight}X >{\RaggedRight}X >{\RaggedRight}X @{}}
        \toprule
        \textbf{Question} & \textbf{Standard Answer} & \textbf{Model Prediction} \\
        \midrule
        
        \multicolumn{3}{@{}l}{\textbf{Case 1: Simple Harmonic Signal Analysis}} \\
        \cmidrule(r){1-3}
        
        What is the type of this signal? & This is a simple harmonic signal. & This is a simple harmonic signal. \\
        
        What is the amplitude of this signal? & The amplitude of this signal is 0.29. & The amplitude of this signal is 0.29. \\
        
        What is the phase of this signal? & The phase of this signal is 2.0 radians. & The phase of this signal is 2.24 radians. \\
        
        What is the base frequency of this signal? & The base frequency of this signal is 50 Hz. & The base frequency is 50 Hz. \\
        
        What is the period of this signal? & The period of this signal is 0.02 seconds. & The period of this signal is 0.14 seconds. \\
        
        What is the peak frequency of this signal? & The peak frequency of this signal is 50.14, which is close to its base frequency. & The peak frequency of this signal is 50 Hz. \\
        
        What is your conclusion? & It represents a single sine wave with a constant amplitude. & The signal oscillates periodically with a single frequency. \\
        \midrule

        \multicolumn{3}{@{}l}{\textbf{Case 2: Single Periodic Impulse Signal Analysis}} \\
        \cmidrule(r){1-3}
        
        What is the type of this signal? & This is a single periodic impulse signal. & This is a single periodic impulse signal. \\
        
        What is the amplitude of this signal? & The amplitude of this signal is 0.68. & The amplitude of this signal is 0.67. \\
        
        What is the base frequency of this signal? & The base frequency is 50 Hz. & The base frequency is 50 Hz. \\
        
        What is the period of this signal? & The period of this signal is 0.02 seconds. & The period of this signal is 0.1 seconds. \\
        
        What is the peak frequency of this signal? & The peak frequency of this signal is 50.14. & The peak frequency of this signal is 50 Hz. \\

        What is the shock interval of this signal? & The shock interval of this signal is [0.12] seconds. & The shock interval of this signal is [0.16] seconds. \\        
        
        What is your conclusion? & This signal has impulse characteristics and decays over time. & This signal decays over time and has impulse characteristics. \\
        
        \bottomrule
    \end{tabularx}
\end{table*}

\begin{table*}[htbp ]
    \centering
    \caption{Examples of model predictions against standard answers for different signal analysis cases (continued from Table \ref{tab:qualitative_comparison_part1}). Note that we only chose the results of three representative signals.}
    \label{tab:qualitative_comparison_part2}
    
    \small 
    
    \renewcommand{\arraystretch}{1.5}       
    
    \begin{tabularx}{\textwidth}{@{} >{\RaggedRight}X >{\RaggedRight}X >{\RaggedRight}X @{}}
        \toprule
        \textbf{Question} & \textbf{Standard Answer} & \textbf{Model Prediction} \\
        \midrule

        \multicolumn{3}{@{}l}{\textbf{Case 3: THU Signal (Outer Fault) Analysis}} \\
        \cmidrule(r){1-3}
        
        What is the type of this signal? & This is a THU signal representing a bearing. & This is a THU signal representing a bearing. \\
        
        What is the fundamental frequency at which this signal was acquired? & The signal was recorded at a fundamental frequency of 10Hz. & The identified characteristic frequency is 10.00 Hz. \\
        
        Which characteristic fault frequency was identified in this signal? & The detected characteristic frequency aligns with the typical fault frequencies associated with outer fault. & The identified characteristic frequency aligns with the typical frequency range of fault frequencies associated with outer fault. \\
        
        How does the identified characteristic frequency correlate with the diagnosed fault? & The identified characteristic frequency is indicative of a outer fault fault in the bearing. & The identified characteristic frequency is 10.00, which is associated with a outer fault. \\
        
        How does the condition of the bearing influence the signal characteristics? & The condition of the bearing influences the signal, allowing for the identification of specific faults like inner, outer, or roller faults. & The condition of the bearing influences the signal characteristics, allowing for the identification of inner, outer, or roller faults. \\
        
        What is your conclusion? & The presence of a characteristic frequency indicates a outer fault. & The presence of a characteristic frequency indicates a outer fault. \\
        
        \bottomrule
    \end{tabularx}
\end{table*}
\clearpage

\section{Evaluation prompt for referee model}

The LLM score is based on GLM-4.1V-Thinking-Flash with human instruction to play the <Vibration Signal Analyst>. The prompt $X_p$ is as follows: 

"Please assess the generated results for the provided vibration signal data. The first result represents the ground truth, and the second is produced by a large language model. Evaluate the second result based on its similarity to the ground truth: the closer the second result is to the ground truth, the higher the score. Additionally, if parameters are identified, assess the accuracy of these parameters, with lower deviation resulting in a higher score. 

Consider the following factors in your evaluation: helpfulness, relevance, accuracy, and expertise. Each factor should contribute to the overall score, with higher similarity across these dimensions resulting in higher scores. 

First, output a single line containing only two scores, separated by a space. The first score should reflect the overall similarity to the ground truth across all criteria, and the second score should reflect the accuracy of parameter identification. Afterward, provide a detailed and unbiased explanation of your evaluation, ensuring that the order of presentation does not influence your judgment. Please remember first output a single line containing only two values indicating the score from 1 to 10, which means only two values appear in the first line without words like score, respectively."



\end{document}